\documentclass[aps,prb,twocolumn,groupedaddress,amsmath,eqsecnum]{revtex4}

    \usepackage{bm}
    \usepackage{graphicx}
\newcommand\beq{\begin{equation}}
\newcommand\eeq{\end{equation}}
\newcommand\beqa{\begin{eqnarray}}
\newcommand\eeqa{\end{eqnarray}}
\newcommand{\nn}{\nonumber\\}
\newcommand{\hs}{\text{HS}}
\newcommand{\fk}{\widetilde{f}}
\newcommand{\hk}{\widetilde{h}}
\newcommand{\ck}{\widetilde{c}}
\newcommand{\wk}{\widetilde{w}}

\newcommand{\dd}{\text{d}}
\newcommand\kk[1]{{{{{#1}}}}}

\begin{document}
\title{Structure  of penetrable-rod fluids: Exact properties and comparison between
Monte Carlo simulations and two analytic theories}
\author{Alexandr Malijevsk\'y}
\email{amail@post.cz}
\affiliation{E. H\'ala Laboratory of
Thermodynamics, Academy of Science of the Czech Republic, Prague 6,
Czech Republic and Institute of Theoretical Physics, Faculty of
Mathematics and Physics, Charles University, Prague 8, Czech
Republic}
\author{Andr\'es Santos}
\email{andres@unex.es}
\homepage{http://www.unex.es/eweb/fisteor/andres/}
\affiliation{Departamento
de F\'{\i}sica, Universidad de Extremadura, E-06071 Badajoz, Spain}
\date{\today}
\begin{abstract}
Bounded potentials are good models to represent the effective
two-body interaction in some colloidal systems, such as dilute
solutions of polymer chains in good solvents. The simplest bounded
potential is that of  penetrable spheres, which takes a positive
finite value if the two spheres are overlapped, being 0 otherwise.
 Even in the
one-dimensional case, the penetrable-rod model is far from trivial,
since interactions are not restricted to nearest neighbors and so
its exact solution is not known. In this paper \kk{the structural
properties of one-dimensional penetrable rods are studied. We} first
derive the exact correlation functions of penetrable-rod fluids to
second order in density at any temperature, as well as in the
high-temperature and zero-temperature limits at any density. \kk{It
is seen that, in contrast to what is generally believed, the
Percus--Yevick equation does not yield the exact cavity function in
the hard-rod limit.} Next, two simple analytic theories are
constructed: a high-temperature approximation based on the exact
asymptotic behavior in the limit $T\to\infty$ and a low-temperature
approximation inspired by the exact result in the opposite limit
$T\to 0$. Finally, we perform Monte Carlo simulations for a wide
range of temperatures and densities to assess the validity of both
theories. It is found that they complement   each other quite well,
exhibiting a good agreement with the simulation data within their
respective domains of applicability and becoming practically
equivalent on the borderline of those domains. \kk{A comparison with
numerical solutions of the Percus--Yevick and the hypernetted-chain
approximations is also carried out. Finally, a} perspective on the
extension of \kk{our two heuristic theories} to the \kk{more
realistic} three-dimensional case is provided.

\end{abstract}
\maketitle

\section{Introduction\label{sec1}}

In the last few years, the study of the structural and thermodynamic
equilibrium properties of fluids with particles interacting via
strongly repulsive potentials have experienced a noticeable revival
because of their interest in the physics of some colloidal systems.
For instance, the effective interaction between two sterically
stabilized colloidal particles can be accurately modeled by the
hard-sphere potential.\cite{L01}

On the other hand, the effective two-body interaction in other
colloidal systems is much softer. For instance, the interaction
potential for star polymers in good solvents can be shown to be
ultrasoft, diverging only logarithmically for short
distances.\cite{L01,LLWAJAR98} Furthermore, in the case of dilute
solutions of polymer chains in good solvents, the centers of mass of
two polymer chains can be separated by a distance smaller  than the
sum of their respective radii of gyration (and even can coincide at
the same point), without violation of the excluded-volume
conditions.\cite{L01} In such a case, the effective two-body
potential is a \textit{bounded} one, being well represented by the
Gaussian core model.\cite{SS97,GL98,LLWL00,LBH00,LLWL01,FHL03}

The simplest bounded potential is that of so-called penetrable
spheres (PS), which is defined as
\beq
\varphi(r)=\left\{
\begin{array}{ll}
\epsilon,& r<\sigma,\\
0,&r>\sigma,
\end{array}
\right.
\label{1}
\eeq
where $\epsilon>0$.  This interaction potential was suggested by
Marquest and Witten\cite{MW89} as a simple theoretical approach to
the explanation of the experimentally observed crystallization of
copolymer mesophases. In the last few years, the PS model has been
the subject of several
studies.\cite{LLWL01,KGRCM94,LWL98,S99,FLL00,RSWL00,SF02,KS02,CG03,AS04,S05}
Density-functional theory\cite{LWL98,S99} predicts a freezing
transition to fcc solid phases with multiply occupied lattice sites.
The existence of clusters of overlapped particles (or ``clumps'') in
the PS crystal and  glass was already pointed out by Klein et
al.,\cite{KGRCM94} who also performed Monte Carlo (MC) simulations
on the system. In the fluid phase, the standard integral equation
theories in general are not very reliable in describing the
structure of the PS fluid, especially inside the core.\cite{LWL98}
While the number of overlapped pairs is  overestimated by the
hypernetted-chain (HNC) theory, it is strongly underestimated by the
Percus--Yevick (PY) theory. Other more sophisticated
closures,\cite{FLL00,CG03} as well as Rosenfeld's
fundamental-measure theory,\cite{RSWL00} are able to predict the
correlations functions with a much higher precision. Only in the
combined high-temperature, high-density limit is the PS model
amenable to an  exact analytical treatment.\cite{AS04} As
applications of the model, let us mention that a mixture of colloids
and non-interacting polymer coils, where the colloid-colloid
interaction is assumed to be that of hard spheres and the
colloid-polymer interaction is described by the PS model, has been
studied.\cite{SF02} The inhomogeneous structure of penetrable
spheres in a spherical pore has also been investigated.\cite{KS02}
In addition, some nonequilibrium properties have been analyzed
recently\cite{S05} and in the past.\cite{GDL79}

As \kk{mentioned} above, the classical integral equation theories
(PY and HNC) do not describe satisfactorily well the structure of
the PS fluid, especially inside the overlapping region, for the
whole range of fluid densities and temperatures. Thus, the PS model
provides a stringent benchmark to test alternative
theories.\cite{FLL00,RSWL00,CG03} Even in the one-dimensional  (1D)
case, the PS model is far from trivial, since interactions are not
restricted to nearest neighbors and so its exact solution is not
known. A surprising consequence of the boundedness of the PS
potential in the 1D case is the plausible existence of a
fluid-crystal phase transition,\cite{AS04} thus providing one of the
rare examples of phase transitions in 1D systems.\cite{CS02}

\kk{Statistical mechanics has a long tradition of studying 1D
systems, especially in those cases where an exact solution to the
many-body problem has been found.\cite{M93} Of course, the 1D PS
model does not intend to describe all the properties of real
polymers in solution, for which spatial dimensionality is known to
be important.\cite{JdC90} However,} it seems worthwhile studying the
1D PS model in order to understand some of the subtleties of the PS
interaction and also to serve as a playground to test theoretical
approaches that can be extended to the more realistic 3D case.

The main aim of this work is to explore the possibility of
constructing simple analytic theories for the structural properties
of the 1D PS fluid, based on known behaviors in the extreme cases of
high and zero temperatures. Before proposing those theories and in
order to gain some insight,
 the exact properties in the limits of low density,
high temperature, and zero temperature are worked out in Sec.\
\ref{sec2}, some technical details being relegated to an Appendix.
\kk{In particular, it is seen that the PY equation does not yield
the exact hard-rod correlation functions for all distances, in
contrast to what is generally believed.\cite{V64,M69}} As a simple
extension to finite temperatures of the  mean-field solution
(asymptotically exact in the combined limit where the reduced
temperature and density go to infinity, its ratio being kept
finite), we propose a high-temperature (HT) approximation in Sec.\
\ref{sec3}. A subtler task consists of the extension to finite
temperatures, imposing some basic continuity conditions, of the
exact solution for hard rods. This is carried out in Sec.\
\ref{sec4}, resulting in what we call the low-temperature (LT)
approximation. Both approximations are compared with our own MC
simulations in Sec.\ \ref{sec5}. It is observed that both theories
complement quite well each other, exhibiting a good agreement with
the simulation data in their respective domains of applicability,
which are wider than what one might have anticipated. To put the
work on a broader perspective, we have also carried out comparisons
with numerical solutions of two integral equation theories, namely
the PY and HNC theories. Both of them fail at low temperatures, but
the HNC theory becomes very accurate and preferable to the PY theory
at moderate and large temperatures. The paper ends in Sec.\
\ref{sec6} with a discussion of the results and a perspective on the
extension of the LT and HT approaches to the 3D case.

\section{Exact properties\label{sec2}}
We consider in this paper a fluid  of  particles on a line
interacting via the pairwise potential (\ref{1}). Henceforth we take
$\sigma=1$ as the length unit and define the reduced temperature as
$T^*\equiv k_BT/\epsilon$. Even in this one-dimensional case, the
exact solution of the problem for arbitrary density and temperature
is not known, the main difficulty lying in the fact that one
particle can interact simultaneously with an arbitrary number of
particles. Thus, it seems convenient to gain some insight by
deriving a few exact results in limiting situations.
\subsection{Low-density limit\label{sec2A}}
Let us introduce the cavity function
\beq
y(r)=e^{\varphi(r)/k_BT}g(r),
\label{2.1}
\eeq
where $g(r)$ is the radial distribution function. In general, $g(r)$
and $y(r)$ depend parametrically on density ($\rho$) and temperature
($T^*$). The virial expansion of the cavity function reads
\beq
y(r)=1+\sum_{n=1}^\infty y_n(r)\rho^n,
\label{2.2}
\eeq
where the coefficients $y_n(r)$ depend parametrically on $T^*$ and
 are represented by diagrams.\cite{B74,HM86} In particular,
\beq
y_1(r)=
\begin{picture}(40,40)(-5,5)
\setlength{\unitlength}{.1mm}
\put(50,100){\circle*{10}}
\put(100,0){\circle{10}}
\put(0,0){\circle{10}}
\put(50,100){\line(1,-2){48}}
\put(50,100){\line(-1,-2){48}}
\end{picture},
\label{2.3}
\eeq
\beq
y_2(r)=
\begin{picture}(40,40)(-5,5)
\setlength{\unitlength}{.1mm}
\put(0,100){\circle*{10}}
\put(100,100){\circle*{10}}
\put(100,0){\circle{10}}
\put(0,0){\circle{10}}
\put(0,100){\line(100,0){100}}
\put(0,100){\line(0,-100){95}}
\put(100,100){\line(0,-100){95}}
\end{picture}
 +2
\begin{picture}(40,40)(-5,5)
\setlength{\unitlength}{.1mm}
\put(0,100){\circle*{10}}
\put(100,100){\circle*{10}}
\put(100,0){\circle{10}}
\put(0,0){\circle{10}}
\put(0,100){\line(100,0){100}}
\put(0,100){\line(0,-100){95}}
\put(100,100){\line(0,-100){95}}
\put(100,100){\line(-1,-1){95}}
\end{picture}+
\frac{1}{2}
\begin{picture}(40,40)(-5,5)
\setlength{\unitlength}{.1mm}
\put(0,100){\circle*{10}}
\put(100,100){\circle*{10}}
\put(100,0){\circle{10}}
\put(0,0){\circle{10}}
\put(0,100){\line(0,-100){95}}
\put(100,100){\line(0,-100){95}}
\put(100,100){\line(-1,-1){95}}
\put(0,100){\line(1,-1){95}}
\end{picture}
+ \frac{1}{2}
\begin{picture}(40,40)(-5,5)
\setlength{\unitlength}{.1mm}
\put(0,100){\circle*{10}}
\put(100,100){\circle*{10}}
\put(100,0){\circle{10}}
\put(0,0){\circle{10}}
\put(0,100){\line(100,0){100}}
\put(0,100){\line(0,-100){95}}
\put(100,100){\line(0,-100){95}}
\put(100,100){\line(-1,-1){95}}
\put(0,100){\line(1,-1){95}}
\end{picture}.
\label{2.4}
\eeq
Here, the open circles represent \textit{root} points separated by a
distance $r$, the filled circles represent \textit{field} points to
be integrated out, and each bond represents a Mayer function
\beq
f(r)=e^{-\varphi(r)/k_BT}-1.
\label{2.5}
\eeq
Thus, for instance,
\beq
\begin{picture}(40,40)(-5,5)
\setlength{\unitlength}{.1mm}
\put(50,100){\circle*{10}}
\put(25,90){3}
\put(100,0){\circle{10}}
\put(75,-10){2}
\put(0,0){\circle{10}}
\put(-25,-10){1}
\put(50,100){\line(1,-2){48}}
\put(50,100){\line(-1,-2){48}}
\end{picture}
=\int \dd\mathbf{r}_3 \,f(r_{13})f(r_{23}),
\label{2.6}
\eeq
\beq
\begin{picture}(40,40)(-5,5)
\setlength{\unitlength}{.1mm}
\put(0,100){\circle*{10}}
\put(-25,90){3} \put(100,100){\circle*{10}}
\put(110,90){4}
\put(100,0){\circle{10}}
\put(75,-10){2}
\put(0,0){\circle{10}}
\put(-25,-10){1}
\put(0,100){\line(100,0){100}}
\put(0,100){\line(0,-100){95}}
\put(100,100){\line(0,-100){95}}
\put(100,100){\line(-1,-1){95}}
\end{picture}=
\int \dd\mathbf{r}_3\int \dd\mathbf{r}_4 \,
f(r_{13})f(r_{34})f(r_{24})f(r_{14}).
\label{2.7}
\eeq

So far, Eqs.\ (\ref{2.1})--(\ref{2.7}) hold for any interaction
potential and any dimensionality. In the special case of penetrable
spheres,  Eq.\ (\ref{2.1}) yields
\beq
g(r)=
\begin{cases}
(1-x)y(r),&r<1,\\
y(r),& r>1,
\end{cases}
\label{2.17}
\eeq
where we have called
\beq
x\equiv 1-e^{-1/T^*}.
\label{2.18}
\eeq
Moreover, the Mayer function becomes
\beq
f(r)=x f_{\text{HS}}(r),
\label{2.8}
\eeq
where
\beq
f_{\text{HS}}(r)=
\begin{cases}
-1,& r<1,\\
0,&r>1,
\end{cases}
\label{2.9}
\eeq
is the Mayer function of hard spheres. Therefore, the spatial
dependence of each one of the diagrams contributing to the virial
expansion (\ref{2.2}) is exactly the same as for hard spheres. The
only difference is that each diagram is now multiplied by the
temperature-dependent parameter $x$ raised to a power equal to the
number of bonds in that particular diagram.

\begin{table}[htb]
\begin{ruledtabular}
\begin{tabular}{ccccc}
Label&Diagram& $0\leq r\leq 1$ & $1\leq r\leq 2$ & $2\leq r\leq 3$\\
\hline
1&$
\begin{picture}(40,40)(-5,5) \setlength{\unitlength}{.1mm}
\put(50,100){\circle*{10}}
\put(100,0){\circle{10}}
\put(0,0){\circle{10}}
\put(50,100){\line(1,-2){48}}
\put(50,100){\line(-1,-2){48}}
\end{picture}$
&$x^2(2-r)$&$x^2(2-r)$&0\\
2A&$
\begin{picture}(40,40)(-5,5) \setlength{\unitlength}{.1mm}
\put(0,100){\circle*{10}}
\put(100,100){\circle*{10}}
\put(100,0){\circle{10}}
\put(0,0){\circle{10}}
\put(0,100){\line(100,0){100}}
\put(0,100){\line(0,-100){95}}
\put(100,100){\line(0,-100){95}}
\end{picture}$
&$-x^{3}\left(3-r^2\right)$&$-\frac{x^{3}}{2}(3-r)^2$&$-\frac{x^{3}}{2}(3-r)^2$\\
2B&$
\begin{picture}(40,40)(-5,5) \setlength{\unitlength}{.1mm}
\put(0,100){\circle*{10}}
\put(100,100){\circle*{10}}
\put(100,0){\circle{10}}
\put(0,0){\circle{10}}
\put(0,100){\line(100,0){100}}
\put(0,100){\line(0,-100){95}}
\put(100,100){\line(0,-100){95}}
\put(100,100){\line(-1,-1){95}}
\end{picture}$
&$\frac{x^{4}}{2}(6-2r-r^2)$&$\frac{x^{4}}{2}(2-r)(4-r)$&0\\
2C&$
\begin{picture}(40,40)(-5,5) \setlength{\unitlength}{.1mm}
\put(0,100){\circle*{10}}
\put(100,100){\circle*{10}}
\put(100,0){\circle{10}}
\put(0,0){\circle{10}}
\put(0,100){\line(0,-100){95}}
\put(100,100){\line(0,-100){95}}
\put(100,100){\line(-1,-1){95}}
\put(0,100){\line(1,-1){95}}
\end{picture}$
&$x^{4}(2-r)^2$&$x^{4}(2-r)^2$&0\\
2D&$
\begin{picture}(40,40)(-5,5) \setlength{\unitlength}{.1mm}
\put(0,100){\circle*{10}}
\put(100,100){\circle*{10}}
\put(100,0){\circle{10}}
\put(0,0){\circle{10}}
\put(0,100){\line(100,0){100}}
\put(0,100){\line(0,-100){95}}
\put(100,100){\line(0,-100){95}}
\put(100,100){\line(-1,-1){95}}
\put(0,100){\line(1,-1){95}}
\end{picture}$
&$-x^{5}(3-2r)$&$-x^{5}(2-r)^2$&0\\
&&&
\end{tabular}
\caption{Explicit expressions of the diagrams contributing to
$y_1(r)$ and $y_2(r)$.}
\label{Table1}
\end{ruledtabular}
\end{table}
In the special case of one-dimensional penetrable spheres (i.e.,
penetrable rods), it is not difficult to evaluate the integrals
represented by the diagrams in Eqs.\ (\ref{2.3}) and (\ref{2.4}).
The results are displayed in Table \ref{Table1}. Note that  diagram
2C is the square of  diagram 1. All the functions are continuous
everywhere and vanish for $r\geq 3$. At $r=1$, diagrams 2A, 2B, and
2D have a second-order discontinuity. At $r=2$, diagrams 1 and 2B
have a first-order discontinuity, while 2C and 2D have a
second-order discontinuity. Finally, 2A has a second-order
discontinuity at $r=3$.

\begin{squeezetable}
\begin{table*}[htb]
\begin{ruledtabular}
\begin{tabular}{cccc}
Function& $0\leq r\leq 1$ & $1\leq r\leq 2$ & $2\leq r\leq 3$\\
\hline
$y_1(r)$&$x^{2}(2-r)$&$x^{2}(2-r)$&$0$\\
 $y_2(r)$
&$-x^{3}\left(3-r^2\right)+x^4\left(8-4r-\frac{1}{2}r^2\right)
-x^5\left(\frac{3}{2}-r\right)$&$-\frac{x^{3}}{2}(r-3)^2+x^4(2-r)\left(5-\frac{3}{2}r\right)
-\frac{x^5}{2}\left(r-2\right)^2$&$-\frac{x^{3}}{2}(r-3)^2$\\
$y_2^{\text{HNC}}(r)$
&$-x^{3}\left(3-r^2\right)+x^4\left(8-4r-\frac{1}{2}r^2\right)
$&$-\frac{x^{3}}{2}(r-3)^2+x^4(2-r)\left(5-\frac{3}{2}r\right)
$&$-\frac{x^{3}}{2}(r-3)^2$\\
$y_2^{\text{PY}}(r)$
&$-x^{3}\left(3-r^2\right)+x^4\left(6-2r-r^2\right)$&$-\frac{x^{3}}{2}(r-3)^2+x^4(2-r)(4-r)$&$-\frac{x^{3}}{2}(r-3)^2$\\
$\lim_{T^*\to 0}y_2(r)$&$\frac{1}{2}r^2-3r+\frac{7}{2}$&$\frac{1}{2}r^2-3r+\frac{7}{2}$&$-\frac{1}{2}(r-3)^2$ \\
$\lim_{T^*\to 0}y_2^{\text{HNC}}(r)$&$\frac{1}{2}r^2-4r+5$&$r^2-5r+\frac{11}{2}$&$-\frac{1}{2}(r-3)^2$ \\
$\lim_{T^*\to
0}y_2^{\text{PY}}(r)$&$3-2r$&$\frac{1}{2}r^2-3r+\frac{7}{2}$&$-\frac{1}{2}(r-3)^2$
\end{tabular}
\caption{Explicit expressions of the functions  $y_1(r)$ and
$y_2(r)$. The first, second, and fifth rows are exact results, while
the third, fourth, sixth, and seventh rows correspond to the HNC and
PY approximations.}
\label{Table2}
\end{ruledtabular}
\end{table*}
\end{squeezetable}
Inserting the expressions displayed in Table \ref{Table1} into Eqs.\
(\ref{2.3}) and (\ref{2.4}) we get the exact results shown in Table
\ref{Table2}. The discontinuities at $r=1$, 2, and 3 are
\beq
y''(1^+)-y''(1^-)=-\rho^2 x^3(3-4x+x^2)+\mathcal{O}(\rho^3),
\label{disc1}
\eeq
\beq
y'(2^+)-y'(2^-)=\rho x^2+\mathcal{O}(\rho^2),
\label{disc2}
\eeq
\beq
y''(3^+)-y''(3^-)=\rho^2 x^3+\mathcal{O}(\rho^3).
\label{disc3}
\eeq
It is interesting to compare the exact density expansion with the
HNC and PY approximations.\cite{B74,HM86} In the HNC approximation,
the so-called  bridge (or elementary) diagrams are neglected. To
second order in density, the only bridge diagram in Table
\ref{Table1} is 2D. Therefore, the function $y_1(r)$ is retained but
the function $y_2(r)$ is approximated by the function
$y_2^{\text{HNC}}(r)$ given in Table \ref{Table2}. In the PY
approximation, apart from the bridge diagrams, a subset of the
remaining diagrams is also neglected. In particular, the PY
expression for $y_2(r)$ neglects
 diagrams 2C and 2D of Table \ref{Table1}, what results in the
function $y_2^{\text{PY}}(r)$ shown in Table \ref{Table2}.

In the hard-rod limit ($\epsilon\to\infty$ or, equivalently, $T^*\to
0$ or $x\to 1$), the exact, HNC, and PY functions $y_2(r)$ reduce to
the expressions also included in Table \ref{Table2}. While the PY
approximation for hard rods yields the exact pair correlation
function for $r\geq 1$ (and this happens to any order in density
because of a mutual cancelation of the neglected diagrams), we see
that the PY cavity function fails inside the overlapping region
($r<1$). In principle, this is not a serious drawback for strict
hard rods, since $g(r)=0$ for $r<1$, regardless of the expression of
the cavity function. However, if we consider penetrable rods at low
temperature ($T^*\ll 1$ or $1-x\ll 1$), then $g(r)$ for $r<1$ takes
on relevant (although small)  non-zero values, so that the
expression of $\lim_{T^*\to 0}y_2^{\text{PY}}(r)$ for $r\leq 1$
 anticipates the sort of limitations one can expect from the solution
of the PY theory to penetrable-sphere models.\cite{FLL00,RSWL00} It
is worth noting that  the second-order discontinuities at $r=1$ of
diagrams 2A, 2B, and 2D of Table \ref{Table1} exactly compensate
each other when those diagrams  are added to get $\lim_{T^*\to
0}y_2(r)$. On the other hand, since  diagram 2D is neglected in the
HNC and PY approximations, an artificial second-order discontinuity
at $r=1$ remains in those approximations. \kk{It is interesting to
point out that, although the hard-rod interaction
($\epsilon\to\infty$) is more singular than the penetrable-rod
interaction ($\epsilon=\text{finite}$), the cavity function $y(r)$
of the former is paradoxically analytic at $r=1$, while that of the
latter presents a second-order discontinuity. When considering the
radial distribution function $g(r)$, however, a jump discontinuity
exists in both cases.}

\begin{figure}[htb]
\includegraphics[width=\columnwidth]{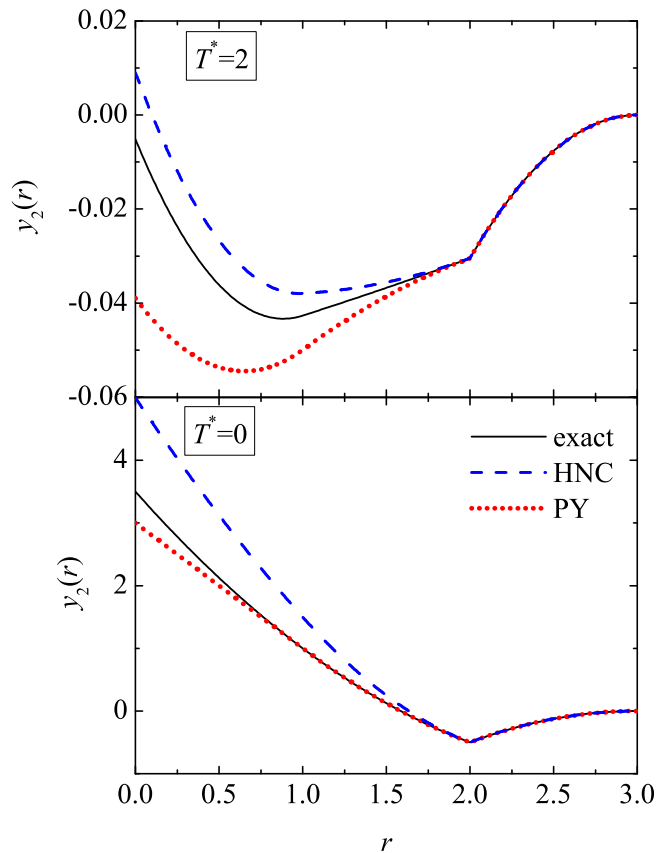}
\caption{(Color online) Plot of the virial coefficient $y_2(r)$ at
$T^*=2$ (top panel) and $T^*=0$ (bottom panel). The solid lines are
the exact results, the dashed lines are the HNC predictions, and the
dotted lines are the PY predictions.
\label{y2}}
\end{figure}
Figure \ref{y2} compares the exact function $y_2(r)$ with the HNC
and PY approximations at $T^*=2$ and $T^*=0$. Although restricted to
low densities and one-dimensional systems, Fig.\ \ref{y2} clearly
illustrates some of the general features found in three-dimensional
penetrable spheres at finite densities:\cite{FLL00,RSWL00} the PY
underestimates the penetrability effect ($r<1$), while the HNC
approximation overestimates it. Moreover, the HNC theory tends to be
better at higher temperatures, while the PY is preferable at lower
temperatures. If we characterize the quality of each approximation
by the separation of the corresponding $y_2(0)$ from the exact
result, it turns out that the temperature beyond which the HNC
approximation becomes better than the PY approximation is $T^*=1/\ln
3\simeq 0.91$. If we take instead $y_2(1)$ as the characteristic
quantity, then the preceding temperature is replaced by $T^*=1/\ln
2\simeq 1.44$. In either case, $T^*\approx 1$ seems to be  the
typical temperature beyond which the HNC approximation prevails over
the PY one. \kk{This will be confirmed later in Sec.\ \ref{sec5}.}

\begin{table*}[htb]
\begin{ruledtabular}
\begin{tabular}{cccccc}
Theory &$B_4(x)$&$B_4(1)$&$T_0^*$&$T^*_{\text{min}}$&$\left.B_4\right|_{\text{min}}$\\
 \hline
Exact&$-x^4\left(2-\frac{7}{2}x+\frac{1}{2}x^2\right)$&$1$&$1.0120$&$1.4427$&$-0.02344$\\
PY-v &$-x^4\left(2-3x\right)$&$1$&$0.9102$&$1.3121$&$-0.03236$\\
PY-c &$-x^4\left(\frac{4}{3}-\frac{7}{3}x\right)$&$1$&$1.1802$&$1.6369$&$-0.01165$\\
PY-e&$-x^4\left(2-\frac{14}{5}x\right)$&$\frac{4}{5}$&$0.7982$&$1.1802$&$-0.04265$\\
HNC-v, HNC-e
&$-x^4\left(2-\frac{7}{2}x\right)$&$\frac{3}{2}$&$1.1802$&$1.6369$&$-0.01747$\\
HNC-c&$-x^4\left(2-\frac{35}{12}x\right)$&$\frac{11}{12}$&$0.8640$&$1.2573$&$-0.03622$
\end{tabular}
\caption{Fourth virial coefficient $B_4(x)$ and other related
quantities as given exactly and by the PY and HNC approximations
through the virial (v), compressibility (c), and energy (e) routes.}
\label{Table2bis}
\end{ruledtabular}
\end{table*}
Although in this paper we are more interested in the structural
properties than in the thermodynamic ones, it is worthwhile
obtaining the first four virial coefficients as functions of
temperature. For penetrable rods, the compressibility factor $Z$,
the isothermal susceptibility $\chi$, and the excess internal energy
per particle $u_{\text{ex}}$ are related to the cavity function as
follows:
\beq
Z\equiv \frac{\beta p}{\rho}=1+\rho x y(1),
\label{2.30}
\eeq
\beqa
\chi\equiv \left(\beta\frac{\partial
p}{\partial\rho}\right)^{-1}&=&1+2\rho\left\{\int_0^1\dd r
[(1-x)y(r)-1]\right.\nn &&\left.+\int_1^\infty \dd r
[y(r)-1]\right\},
\label{chi}
\eeqa
\beq
u_{\text{ex}}=\rho\epsilon(1-x)\int_0^1\dd r y(r),
\label{uex}
\eeq
where $p$ is the pressure and $\beta\equiv 1/k_BT$. The virial
coefficients $B_n(x)$ are defined by the density expansion
\beq
Z=1+\sum_{n=2}^\infty B_n(x)\rho^{n-1}.
\label{virial}
\eeq
Inserting the exact cavity function to second order in density into
Eqs.\ (\ref{2.30})--(\ref{uex}), and taking into account the
thermodynamic relations $\chi^{-1}=\partial(\rho Z)/\partial\rho$
and $\rho\partial u_{\text{ex}}/\partial\rho=\partial
Z/\partial\beta$, one consistently gets $B_2(x)=x$, $B_3(x)=x^3$,
and the fourth virial coefficient $B_4(x)$ displayed in Table
\ref{Table2bis}. On the other hand, insertion of $y^{\text{PY}}(r)$
yields a different $B_4(x)$ depending on whether the virial (v)
route (\ref{2.30}), the compressibility (c) route (\ref{chi}), or
the energy (e) route (\ref{uex}) is used. This internal
inconsistency is also present in the case of $y^{\text{HNC}}(r)$,
except that now the virial and energy routes are
equivalent.\cite{BH76} These approximate expressions for the fourth
virial coefficient are also included in Table \ref{Table2bis}. The
exact and the approximate $B_4(x)$ share the properties of changing
sign at a certain temperature $T_0^*$ and having a (negative)
minimum value $\left.B_4\right|_{\text{min}}$ at a higher
temperature $T^*_{\text{min}}$. Only the PY-v and PY-c
approximations are consistent with the exact hard-rod value
$B_4(1)=1$, while only the PY-c approximation fails to yield the
correct limit $B_4\to -2 x^4$ for high temperatures. {}From Table
\ref{Table2bis} one can conclude that the best global agreement with
the exact $B_4(x)$ is achieved by the PY-v approximation. It is
interesting to note that, while the energy route to the equation of
state is ill defined for hard spheres, one can circumvent this
problem by first getting $B_4(x)$ through the energy route and then
taking the zero-temperature limit $\lim_{x\to 1}B_4(x)$. The fact
that $B_4^{\text{PY-e}}(1)\neq B_4^{\text{PY-v}}(1)$ shows that the
equivalence between the energy and virial routes proven in Ref.\
\onlinecite{S05b} when hard spheres are obtained from the
square-shoulder potential does not hold when hard spheres are
obtained from penetrable spheres.

\subsection{High-temperature limit\label{sec2B}}
In the high-temperature limit ($T^*\to\infty$), the parameter
defined by Eq.\ (\ref{2.18}) tends to zero, i.e., $x\approx 1/T^*\to
0$. Since a diagram having $m$ bonds is of order $x^m$, we can
neglect all the diagrams contributing to a given coefficient
$y_n(r)$, except the one having the least number of bonds (namely,
$m=n+1$ bonds). In other words, only the linear chain diagrams
survive:\cite{AS04}
\beqa
y_n(r_{12})&\to &\;
\stackrel{1}{\circ}\!\!\text{---}\!\!\stackrel{3}{\bullet}\!\!\text{---}\!\!\stackrel{4}{\bullet}\!\!\text{---}\cdots
\text{---}\!\!\!\!\!\stackrel{n+2}{\bullet}\!\!\!\!\!\text{---}\!\!\stackrel{2}{\circ}
\nn&=&x^{n+1}\int \dd\mathbf{r}_3\int \dd\mathbf{r}_4\ldots\int
\dd\mathbf{r}_{n+2} \nn &&\times
f_{\text{HS}}(r_{13})f_{\text{HS}}(r_{34})\cdots
f_{\text{HS}}(r_{n+2,2}).
\label{linear}
\eeqa
In Fourier space,
\beq
\widetilde{y}_n(k)\to x^{n+1}\left[\fk_\hs(k)\right]^{n+1},
\label{2.19}
\eeq
where $\fk_\hs(k)$ is the Fourier transform of the hard-sphere Mayer
function (\ref{2.9}). {}From Eqs.\ (\ref{2.2}) and (\ref{2.19}) it
is straightforward to get the Fourier transform of $y(r)-1$, so that
we finally have
\beq
y(r)\to 1+ x w(r),
\label{2}
\eeq
where
 $w(r)$ is the inverse Fourier transform of
\beq
\widetilde{w}(k)={\rho x}\frac{[\widetilde{f}_\hs(k)]^2}{1-{\rho}
x\widetilde{f}_\hs(k)}.
\label{4.1b}
\eeq

The limit result (\ref{2}) can be written in a number of equivalent
ways. For instance, the total correlation function $h(r)\equiv
g(r)-1$ becomes, on account of Eqs.\ (\ref{2.17}), (\ref{2.9}), and
(\ref{2}),
\beq
h(r)\to x\left[w(r)+f_\hs(r)\right].
\label{2.20}
\eeq
In Fourier space,
\beq
\hk(k)\to x\frac{\fk_\hs(k)}{1-\rho x \fk_\hs(k)},
\label{2.21}
\eeq
where use has been made of Eq.\ (\ref{4.1b}). {}From Eq.\
(\ref{2.21}) it is straightforward to get the structure factor
$S(k)=1+\rho\hk(k)$, as well as the Fourier transform
$\ck(k)=\hk(k)/S(k)$:
\beq
S(k)\to \frac{1}{1-\rho x\fk_\hs(k)},\quad \ck(k)\to x\fk_\hs(k).
\label{2.22}
\eeq
The last expression in Eq.\ (\ref{2.22}) is equivalent to $c(r)\to
f(r)$.

The asymptotic behaviors (\ref{2})--(\ref{2.22}) are of mean-field
type\cite{LLWL01,G75,GK77,KG80,KB81} and hold in the combined limit
$T^*\to\infty$, $\rho\to\infty$ with $\rho
x\approx\rho/T^*=\text{fixed}$ for any dimensionality. In the
one-dimensional case, one simply has
\beq
\widetilde{f}_\hs(k)=-2\sin k/k,
\label{2.23}
\eeq
so that
\beq
w(r)=\frac{1}{\pi}\int_0^\infty \dd k\, \cos kr \,\widetilde{w}(k),
\quad \widetilde{w}(k)=\frac{4{\rho}x\sin^2k}{k^2+2{\rho}x k\sin k}.
\label{4}
\eeq
 It is proven in the Appendix  that the density expansion of
$w(r)$ is
\beq
w(r)=\sum_{n=2}^\infty (\rho x)^{n-1} w_n(r),
\label{2.24}
\eeq
where
\beq
w_n(r)=n\sum_{m=0}^n
\frac{(-1)^{n+m}}{m!(n-m)!}(n-2m-r)^{n-1}\Theta(n-2m-r),
\label{2.25}
\eeq
$\Theta(x)$ being the Heaviside step function. It is also shown that
\beq
2\frac{w_n(1)}{n}=-\frac{w_{n+1}(0)}{n+1},\quad n\geq 2,
\label{2.29}
\eeq
this property being needed to prove the thermodynamic consistency
between the virial and energy routes to the equation of
state.\cite{AS04}

The virial series (\ref{2.24}) converges for $\rho x<\nu(r)$, where
$1/\nu(r)=\lim_{n\to\infty}|w_{n+1}(r)/w_{n}(r)|$. Equation
(\ref{2.29}) implies that $\nu(0)=\nu(1)$, but otherwise the radius
of convergence $\nu(r)$ could in principle be $r$-dependent. We have
numerically checked that this is not the case and that
$\nu(r)=\frac{1}{2}$ regardless of the value of $r$. \kk{{}From Eq.\
(\ref{4}) we see that the radius of convergence $\nu(r)=\frac{1}{2}$
is a consequence of the mathematical singularity of $w(r)$ at the
negative density $2\rho x=-1$. Therefore, we conclude that in the
high-temperature domain the virial expansion of the cavity function
converges uniformly for $\rho<1/2x\approx T^*/2$.}

 Using Eqs.\ (\ref{2}) and (\ref{2.24}) in
Eq.\ (\ref{2.30}), we obtain the virial expansion of the equation of
state in the high-temperature limit,
\beq
\frac{p}{\rho k_BT}\to 1+\rho x+\sum_{n=3}^\infty \rho^{n-1}x^n
w_{n-1}(1),
\label{2.31}
\eeq
so that the virial coefficients are $B_2(x)=x$ and  $B_n(x)\to x^n
w_{n-1}(1)$ for $n\geq 3$. Therefore, the virial series of the
equation of state converges for $\rho<(2x)^{-1}$ in the
high-temperature limit.

By an adequate reordering of terms (see the Appendix), the function
$w(r)$ can be written in real space as
\beq
w(r)=\Theta(1-r)+2\sum_{p=1}^\infty (-1)^p J_{p}'(2\rho
x(p-r))\Theta(p-r),
\label{2.26}
\eeq
where $J_p'(z)$ is the first derivative of the Bessel function of
the first kind $J_p(z)$.\cite{AS72} In particular, in the
overlapping region ($r<1$), we have
\beq
w(r<1)=1+2\sum_{p=1}^\infty (-1)^p J_{p}'(2\rho x(p-r)).
\label{2.27}
\eeq
Likewise, in the shell $n<r<n+1$ with $n\geq 1$,
\beq
w(n<r<n+1)=2\sum_{p=n+1}^\infty (-1)^p J_{p}'(2\rho x(p-r)).
\label{2.28}
\eeq
Using known properties of the Bessel functions it is possible to
derive some exact results. For instance, the second-order,
first-order, and second-order discontinuities of $w(r)$ at $r=1$, 2,
and 3, respectively, are
\beq
w''(1^+)-w''(1^-)=-3(\rho x)^2,
\label{d1}
\eeq
\beq
w'(2^+)-w'(2^-)=\rho x,
\label{d2}
\eeq
\beq
w''(3^+)-w''(3^-)=(\rho x)^2.
\label{d3}
\eeq
Equations (\ref{d1})--(\ref{d3}) give the singularities of the
cavity function for any density in the high-temperature limit. Of
course, they are compatible with Eqs. (\ref{disc1})--(\ref{disc3})
for low densities. Other interesting results are
\beq
w'(1)=\rho x\left[w(0)-w(2)-1\right],
\label{d4}
\eeq
\beq
w'(2^+)=\rho x\left[w(1)-w(3)\right],
\label{d5}
\eeq
\beq
w'(n)=\rho x\left[w(n-1)-w(n+1)\right], \quad n\geq 3.
\label{d6}
\eeq

\begin{figure}[htb]
\includegraphics[width=\columnwidth]{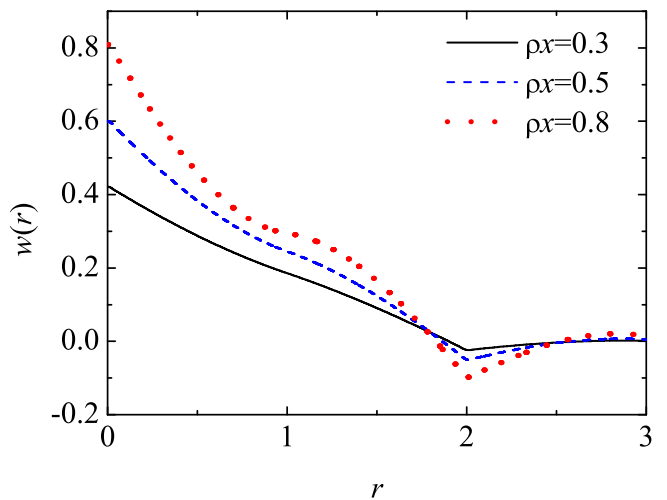}
\caption{(Color online) Plot of the function $w(r)$
 at $\rho x=0.3$, 0.5, and 0.8.
\label{wr}}
\end{figure}
Although the series representation (\ref{2.26}) is useful to derive
explicit results and seems to converge for any value of $\rho x$, it
is more practical to use the integral representation (\ref{4}) from
a computational point of view.
 Figure \ref{wr} shows the function
$w(r)$ for $\rho x=0.3$, $\rho x=0.5$, and $\rho x=0.8$. As
expected, the function $w(r)$ becomes more structured as the product
$\rho x$ grows.

\subsection{Zero-temperature limit: Hard rods\label{sec2C}}
The correlation functions for one-dimensional fluids with
interactions restricted to nearest neighbors  are exactly
known.\cite{M69,SZK53,PL54,K55,LZ71,HC04} However, as \kk{indicated}
in Sec.\ \ref{sec1}, the bounded nature of the potential (\ref{1})
allows for multiple pair interactions and, as a consequence,
 the interactions are obviously not restricted to
nearest neighbors.

On the other hand, as the temperature is lowered, the occurrence of
simultaneous overlapping of three or more particles becomes less and
less likely and the nearest-neighbor approximation becomes more and
more accurate, until the exact result is obtained at zero
temperature (hard-rod limit). In that limit ($T^*=0$ or $x=1$), the
radial distribution function $g(r)$ vanishes inside the core ($r<1$)
but the cavity function $y(r)$ does not and so special care must be
taken to get $\lim_{T^*\to 0} y(r)$ as a function of density for
$r<1$.  As already recognized by Stell,\cite{S63} in order to get
the hard-sphere cavity function for $r<1$ one must take the limit of
a suitably parameterized non-singular potential.

We start by recalling the exact solution in the case of
nearest-neighbor interactions. At arbitrary density and temperature,
the one-dimensional radial distribution function can be written
as\cite{LZ71,HC04}
\beq
g(r)=\frac{1}{\rho}\sum_{n=1}^\infty p_n(r),
\label{5}
\eeq
where $p_n(r)$ is the probability density distribution that the
$n$-th neighbor of a given particle is a distance $r$ apart. The
distributions $\{p_n(r)\}$ obey the recursive relation
\beq
p_n(r)=\int_0^r\dd r'\, p_{n-1}(r')p_1(r-r'),\quad n\geq 2.
\label{6}
\eeq
The explicit form of the nearest-neighbor probability distribution
is
\beq
p_1(r)=K e^{-\varphi(r)/k_BT} e^{-\xi r},
\label{2.32}
\eeq
where the amplitude $K$ and the damping constant $\xi$ will be
determined later on.

The convolution in (\ref{6}) suggests the introduction of the
Laplace transforms
\beq
P_n(t)=\int_0^\infty \dd r\, e^{-rt}p_n(r),\quad G(t)=\int_0^\infty
\dd r\, e^{-rt}g(r).
\label{7}
\eeq
Thus, Eq.\ (\ref{6}) yields
\beq
P_n(t)=P_{n-1}(t)P_1(t)=P_1^n(t),\quad n\geq 2.
\label{8}
\eeq
When this is introduced into Eq.\ (\ref{5}) one gets
\beq
G(t)=\frac{1}{\rho}\sum_{n=1}^\infty
[P_1(t)]^n=\frac{1}{\rho}\frac{P_1(t)}{1- P_1(t)}.
\label{9}
\eeq
{}From Eq.\ (\ref{2.32}), it is obtained
\beq
P_1(t)=K\Omega(t+\xi),
\label{2.33}
\eeq
where
\beq
\Omega(t)=\int_0^\infty \dd r\, e^{-rt} e^{-\varphi(r)/k_BT}
\label{2.34}
\eeq
is the Laplace transform of $\exp[-\varphi(r)/k_BT]$.

To close the solution, one needs to determine the parameters $K$ and
$\xi$. This is done by imposing basic consistency conditions. The
relationship between $G(t)$ and the Laplace transform $H(t)$ of the
total correlation function $h(r)=g(r)-1$ is
\beq
G(t)=\frac{1}{t}+H(t).
\label{16}
\eeq
Since $H(t)$ must be finite at $t=0$ (from the compressibility route
to the equation of state), it follows that
\beq
G(t)=\frac{1}{t}+\mathcal{O}(t^0),\quad
P_1(t)=1-\frac{1}{\rho}t+\mathcal{O}(t^2).
\label{17}
\eeq
The last equality in (\ref{17}) implies that $P_1(0)=1$ and
$P_1'(0)=-1/\rho$. Hence, $K=1/\Omega(\xi)$ and $\xi$ is the
solution of the equation
\beq
\frac{\Omega'(\xi)}{\Omega(\xi)}=-\frac{1}{\rho}.
\label{2.35}
\eeq

We emphasize that Eqs.\ (\ref{5})--(\ref{2.35}) provide the exact
solution only in the case of nearest-neighbor interactions. Let us
now ``forget'' for the moment that the penetrable-sphere potential
(\ref{1}) is not restricted to nearest neighbors and apply the above
scheme to it. In that approximation, Eqs.\ (\ref{2.32}) and
(\ref{2.34}) yield
\beq
p_1(r)=K e^{-\xi r}\left[1-x+x\Theta(r-1)\right],
\label{2.36}
\eeq
\beq
\Omega(t)=\frac{1-x+x e^{-t}}{t}.
\label{2.37}
\eeq
The damping coefficient $\xi$  is the solution of the transcendental
equation (\ref{2.35}), i.e.,
\beq
\frac{x}{1-x}\left(\xi-\xi'\right) e^{-\xi}=\xi',
\label{2.38}
\eeq
where we have called
\beq
\xi'\equiv \frac{\xi}{\rho}-1.
\label{2.40}
\eeq
In addition, the amplitude $K$ becomes
\beq
K=\frac{\xi'e^{\xi}}{x }
\label{2.39b}
\eeq
and Eq.\ (\ref{2.33}) gives
\beq
P_1(t)=\frac{\xi-\xi'+\xi' e^{-t}}{t+\xi}.
\label{2.39}
\eeq
Inserting this into (\ref{9}) and expanding in powers of $e^{-t}$,
we get
\beqa
\rho G(t)&=&\frac{\xi-\xi'}{t+\xi'}+\sum_{n=1}^\infty
{\xi'}^n\left[\frac{1}{(t+\xi')^n}+\frac{\xi-\xi'}{(t+\xi')^{n+1}}\right]e^{-nt}.\nn
\label{2.41}
\eeqa
Its Laplace inversion yields
\beq
g(r)=\sum_{n=0}^\infty \psi_n(r-n)\Theta(r-n),
\label{20}
\eeq
where
\beq
\psi_n(r)=\frac{{\xi'}^{n}}{\rho}\frac{e^{-\xi'
r}r^{n-1}}{n!}\left[n+(\xi-\xi')r\right].
\label{21}
\eeq
In particular, using Eq.\ (\ref{2.17}), the cavity function inside
the core is, in this nearest-neighbor approximation,
\beq
y(r<1)=\frac{\psi_0(r)}{1-x}=\frac{\xi-\xi'}{\rho(1-x)}e^{-\xi' r}.
\label{2.42}
\eeq

Of course, Eqs.\ (\ref{2.36})--(\ref{2.42}) are not exact at finite
temperature. However, they become exact in the hard-rod limit
($T^*\to 0$ or $x\to 1$). In that limit, the solution of Eq.\
(\ref{2.38}) is $\xi=\xi_0-\xi_0^2 e^{\xi_0}(1-x)+\cdots$, so that
$\xi'=\xi-\xi_0 e^{\xi_0}(1-x)+\cdots$, where
\beq
\xi_0\equiv \frac{\rho}{1-\rho}.
\label{2.43}
\eeq
Therefore,
\beq
\lim_{T^*\to 0}\psi_n(r)=\frac{\xi_0^n}{\rho (n-1)!}e^{-\xi_0
r}r^{n-1},\quad n\geq 1,
\label{2.44}
\eeq
\beq
\lim_{T^*\to 0}y(r<1)= \frac{\xi_0}{\rho}e^{-\xi_0(r-1)}.
\label{2.45}
\eeq
The \kk{final} explicit expression of the cavity function for hard
rods is then
\beqa
\lim_{T^*\to 0}y(r)&=& \frac{\xi_0}{\rho}e^{-\xi_0(r-1)}+
\sum_{n=2}^\infty \frac{\xi_0^n}{\rho (n-1)!}e^{-\xi_0 (r-n)}\nn
&&\times
(r-n)^{n-1}\Theta(r-n).
\label{2.46}
\eeqa
This equation not only gives  the exact hard-rod radial distribution
function $g(r)=y(r)\Theta(r-1)$, but also the cavity function inside
the hard core ($r<1$). Interestingly enough, $y(r)$ for $r<1$ is
just the analytical continuation of its expression for $1<r<2$, so
that $y(r)$ is analytical at $r=1$, even though the potential is
highly singular at that point. This singularity manifests itself at
$r=n$ with $n\geq 2$, where $y(r)=g(r)$ presents a discontinuity of
order $n-1$.

\kk{We are not aware of any previous derivation of Eq.\
(\ref{2.45}).} By expanding in powers of density, it is easy to
verify that Eq.\ (\ref{2.46}) is consistent with the exact
$\lim_{T^*\to 0}y_1(r)$ and  $\lim_{T^*\to 0}y_2(r)$ given in Table
\ref{Table2}. As a further test of the exact character of Eq.\
(\ref{2.45}), let us check the fulfillment of the following two
zero-separation theorems,\cite{L95} as applied to hard rods,
\beq
\ln y(0)=\frac{1}{k_BT}\mu_{\text{ex}},
\label{36}
\eeq
\beq
\left.\frac{\partial \ln y(r)}{\partial r}\right|_{r=0}=-\rho y(1).
\label{37}
\eeq
In Eq.\ (\ref{36}), $\mu_{\text{ex}}$ is the excess chemical
potential, which is given by
\beq
\frac{1}{k_BT}\mu_{\text{ex}}=Z(\rho)-1+\int_0^\rho
\dd\rho'\frac{Z(\rho')-1}{\rho'},
\label{38}
\eeq
where $Z=1+\rho y(1)$ is the hard-rod compressibility factor [cf.\
Eq.\ (\ref{2.30})]. Since $y(1)=1/(1-\rho)$, Eqs.\ (\ref{36}) and
(\ref{37}) imply that
\beq
y(0)=\frac{e^{\rho/(1-\rho)}}{1-\rho}=y(1)e^{\rho y(1)},
\label{39}
\eeq
\beq
\left.\frac{\partial \ln y(r)}{\partial
r}\right|_{r=0}=-\frac{\rho}{1-\rho}.
\label{40}
\eeq
It is straightforward to check that Eq.\ (\ref{2.45}) verifies both
(\ref{39}) and (\ref{40}).

In a general one-dimensional system, the Fourier transforms of the
total and direct correlation functions can be obtained from the
Laplace transform of the radial distribution function as
\beq
\hk(k)=G(t=ik)+G(t=-ik),\quad \ck(k)=\frac{\hk(k)}{1+\rho\hk(k)}.
\label{2.47}
\eeq
Inserting Eq.\ (\ref{2.39}) into Eq.\ (\ref{9}) and taking the
zero-temperature limit one finally gets
\beq
\lim_{T^*\to 0}\ck(k)=-\frac{2\xi_0}{\rho}\left[\xi_0\frac{1-\cos
k}{k^2}+\frac{\sin k}{k}\right],
\label{2.48}
\eeq
\beq
\lim_{T^*\to 0} c(r)=-\frac{1-\rho r}{(1-\rho)^2}\Theta(1-r).
\label{2.49}
\eeq
The bridge function is defined as\cite{HM86} $B(r)=\ln
y(r)+c(r)-h(r)$. Therefore, Eqs.\ (\ref{2.45}) and (\ref{2.49})
yield
\beq
\lim_{T^*\to 0} B(r<1)=-\frac{\rho(1-\rho
r)}{(1-\rho)^2}-\ln(1-\rho).
\label{2.51}
\eeq

As is well known, the PY theory yields the exact correlation
functions $g(r)$ and $c(r)$ for hard rods.\cite{V64,M69,S63}
However, it provides a wrong cavity function $y(r)$ inside the hard
core, as already seen to order $\rho^2$ in Sec.\ \ref{sec2A}. To
show this for finite densities, we recall that the PY approximation
consists of the closure\cite{HM86} $y(r)=g(r)-c(r)$. Therefore,
\beq
\lim_{T^*\to 0}y^{\text{PY}}(r<1)= -c(r<1)=\frac{1-\rho
r}{(1-\rho)^2}.
\label{2.50}
\eeq
\kk{This result, which gives rise to a second-order discontinuity at
$r=1$, differs markedly from} the exact result (\ref{2.45}). The PY
cavity function $\lim_{T^*\to 0}y^{\text{PY}}(r)$ verifies the
zero-separation theorem (\ref{40}) but not (\ref{39}). As for the PY
bridge function inside the core, the result is
\beq
\lim_{T^*\to 0} B^{\text{PY}}(r<1)=-\frac{\rho(2-r-\rho
)}{(1-\rho)^2}+\ln\frac{1-\rho r}{(1-\rho)^2},
\label{2.53}
\eeq
which differs from Eq.\ (\ref{2.51}).

\begin{figure}[htb]
\includegraphics[width=\columnwidth]{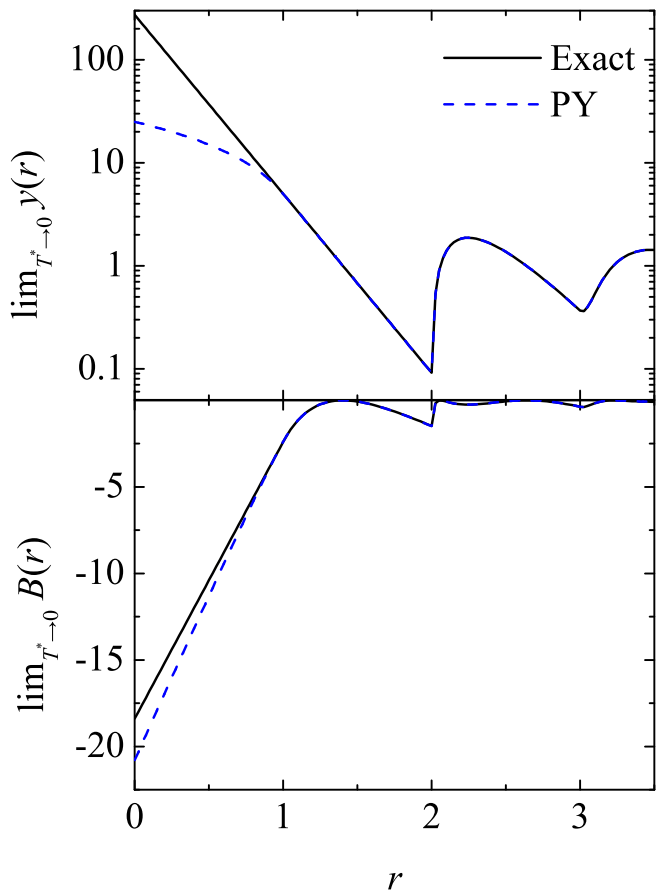}
\caption{(Color online) Plot of the cavity function $y(r)$ and the
bridge function $B(r)$ in the hard-rod limit at $\rho=0.8$. The
solid lines are the exact results, while the dashed lines are the PY
predictions.
\label{yHS}}
\end{figure}
Figure \ref{yHS} shows $\lim_{T^*\to 0}y(r)$ and $\lim_{T^*\to
0}B(r)$ at $\rho=0.8$. It clearly illustrates the tendency of the PY
approximation to underestimate the cavity function inside the core,
a feature that continues to hold in the case of three-dimensional
penetrable spheres.\cite{FLL00} It is also quite apparent that the
HNC closure $B(r)=0$ is far from being accurate in the hard-rod
limit, especially for $r<1$.

\kk{Before closing this Section it is worthwhile noting a flaw in
the ``proof'' given in Ref.\ \onlinecite{M69} on the exact character
of Eq.\ (\ref{2.50}). It is first proven that the PY closure is
consistent with the exact hard-core radial distribution function or,
equivalently, with the exact cavity function for $r>1$. Next, taking
the limit $r\to 1^+$ and taking into account the continuity of
$y(r)$  at $r=1$, it is argued that
\beqa
\frac{1-\rho r}{(1-\rho)^2}&=&\lim_{r\to 1^+}\lim_{T^*\to
0}y(r)=\lim_{r\to 1^-}\lim_{T^*\to 0}y(r)\nn &=&\lim_{T^*\to
0}y(r<1).
\label{2.50bis}
\eeqa
While the two first equalities are entirely correct, the third one
is not justified in general, so that the conclusion that
$\lim_{T^*\to 0}y(r<1)=\lim_{T^*\to 0}y^{\text{PY}}(r<1)$ is flawed.
 }

\section{High-temperature approximation\label{sec3}}
In the limit of asymptotically high temperatures, the cavity
function becomes the mean-field result (\ref{2}), where the function
$w(r)$, which in the one-dimensional case is given by Eqs.\
(\ref{4}) or (\ref{2.26}), depends on density and temperature
through the product $\rho x$ only. It is then natural to expect that
the simple ansatz
\beq
y(r)=1+x w(r)
\label{3.0}
\eeq
constitutes a good approximation for sufficiently large
temperatures. As a matter of fact, comparison with three-dimensional
simulation results shows that this indeed the case for $T^*>
3$.\cite{LLWL01}

The approximation (\ref{3.0}) neglects all the diagrams different
from the linear chain ones. A better approximation can be expected
by making
\beq
y(r)=\exp[x w(r)],
\label{3.1}
\eeq
which reduces to (\ref{2}) in the limit $x\to 0$. In addition to the
linear chain diagrams, the approximation (\ref{3.1}) retains all the
reducible diagrams that factorize into linear chain diagrams. In
particular, (\ref{3.1}) implies that
\beq
y_2(r)=
\begin{picture}(40,40)(-5,5)
\setlength{\unitlength}{.1mm}
\put(0,100){\circle*{10}}
\put(100,100){\circle*{10}}
\put(100,0){\circle{10}}
\put(0,0){\circle{10}}
\put(0,100){\line(100,0){100}}
\put(0,100){\line(0,-100){95}}
\put(100,100){\line(0,-100){95}}
\end{picture}
+ \frac{1}{2}
\begin{picture}(40,40)(-5,5)
\setlength{\unitlength}{.1mm}
\put(0,100){\circle*{10}}
\put(100,100){\circle*{10}}
\put(100,0){\circle{10}}
\put(0,0){\circle{10}}
\put(0,100){\line(0,-100){95}}
\put(100,100){\line(0,-100){95}}
\put(100,100){\line(-1,-1){95}}
\put(0,100){\line(1,-1){95}}
\end{picture}
\label{3.2},
\eeq
\beq
y_3(r)=
\begin{picture}(40,40)(-5,10)
\setlength{\unitlength}{.1mm}
\put(0,100){\circle*{10}}
\put(50,150){\circle*{10}}
\put(100,100){\circle*{10}}
\put(100,0){\circle{10}}
\put(0,0){\circle{10}}
\put(0,100){\line(1,1){50}}
\put(100,100){\line(-1,1){50}}
\put(0,100){\line(0,-100){95}}
\put(100,100){\line(0,-100){95}}
\end{picture}
+\begin{picture}(40,40)(-5,10) \setlength{\unitlength}{.1mm}
\put(0,100){\circle*{10}}
\put(100,100){\circle*{10}}
\put(50,150){\circle*{10}}
\put(100,0){\circle{10}}
\put(0,0){\circle{10}}
\put(0,100){\line(100,0){100}}
\put(0,100){\line(0,-100){95}}
\put(100,100){\line(0,-100){95}}
\put(50,150){\line(-1,-3){48}}
\put(50,150){\line(1,-3){48}}
\end{picture}
+\frac{1}{6}
\begin{picture}(40,40)(-5,10)
\setlength{\unitlength}{.1mm}
\put(0,100){\circle*{10}}
\put(100,100){\circle*{10}}
\put(50,150){\circle*{10}}
\put(100,0){\circle{10}}
\put(0,0){\circle{10}}
\put(0,100){\line(0,-100){95}}
\put(100,100){\line(0,-100){95}}
\put(100,100){\line(-1,-1){95}}
\put(0,100){\line(1,-1){95}}
\put(50,150){\line(-1,-3){48}}
\put(50,150){\line(1,-3){48}}
\end{picture}
\label{3.3}.
\eeq

A third possibility is the Pad\'e approximant
\beq
y(r)=\frac{1}{1-x w(r)}.
\label{3.4}
\eeq
In that case, the diagrams retained are the same as in the
approximation (\ref{3.1}), except that the numerical factors in
front of the reducible diagrams are larger. For instance, the
factors affecting the second diagram of Eq.\ (\ref{3.2}) and the
second and third diagrams of Eq.\ (\ref{3.3}) are 1 (instead of
$\frac{1}{2}$), 2 (instead of 1), and 1 (instead of $\frac{1}{6}$),
respectively. This can be interpreted as an effective way of
compensating for the neglected diagrams.

\begin{figure}[htb]
\includegraphics[width=\columnwidth]{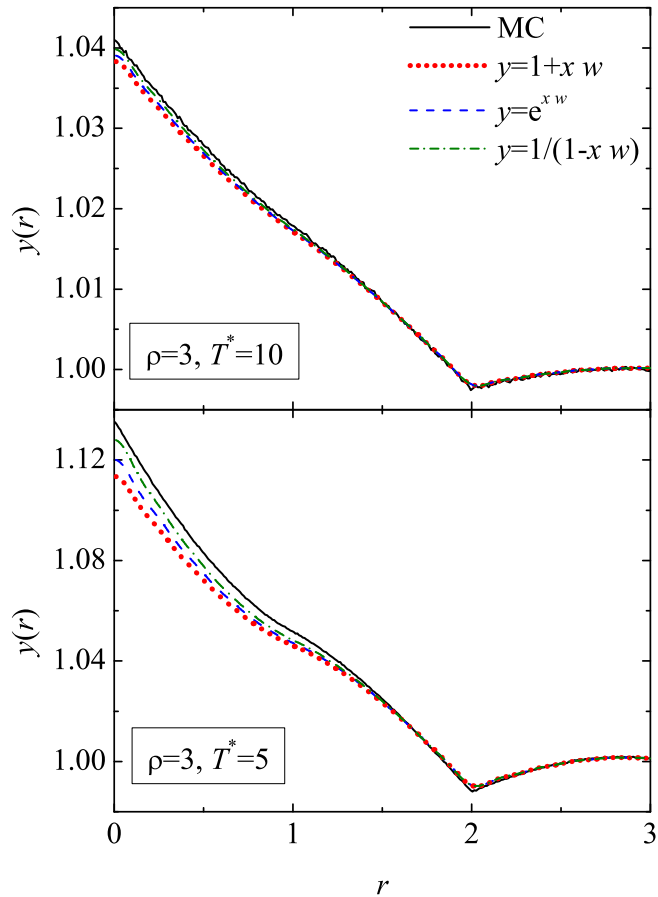}
\caption{(Color online) Plot of the cavity function $y(r)$ at
$\rho=3$ and $T^*=10$ (top panel) and $T^*=5$ (bottom panel). The
solid lines are results obtained from MC simulations, the dotted
lines represent the approximation (\ref{3.0}), the dashed lines
represent the approximation (\ref{3.1}), and the dashed-dotted lines
represent the approximation (\ref{3.4}).
\label{yT5&10}}
\end{figure}
The three heuristic approximations (\ref{3.0}), (\ref{3.1}), and
(\ref{3.4}), among other possible ones, have the potential of
describing satisfactorily well the structure of penetrable-rod
fluids at sufficiently high temperatures. To test this expectation
and also to determine which approximation is the most accurate one,
we have compared them with MC simulations at $T^*=10$ and $T^*=5$ in
Fig.\ \ref{yT5&10}. It can be observed that the three approximations
tend to underestimate the correlation function for $r<2$, this
effect becoming more pronounced as the temperature decreases. On the
other hand, the Pad\'e approximant (\ref{3.4}) is closer to the MC
curves than the exponential approximation (\ref{3.1}) and the linear
approximation (\ref{3.0}). Therefore, in what remains we will adopt
Eq.\ (\ref{3.4}), \kk{supplemented by Eq.\ (\ref{4}),} as our
high-temperature (HT) approximation. \kk{The highest value of $w(r)$
occurs at $r=0$ and is an increasing function of $\rho x$. For $\rho
x\leq 1.1$ one has $w(0)\leq 1$ and so Eq.\ (\ref{3.4}) is
mathematically well defined for any temperature if $\rho x\leq 1.1$.
This is the range where the fluid is expected to be stable against
the solid.\cite{AS04}}

\section{Low-temperature approximation\label{sec4}}

\subsection{Construction of the approximation}
We have already mentioned that the automatic translation to
penetrable rods [cf.\ Eqs.\ (\ref{2.36})--(\ref{2.42})] of the
scheme valid for nearest-neighbor interactions does not yield the
exact correlation functions, except at zero temperature. On the
other hand, one might reasonably wonder whether Eqs.\
(\ref{2.36})--(\ref{2.42}) constitute a good approximation for
(non-zero) low temperatures (i.e., $T^*\ll 1$ or $1-x\ll 1$). A
drawback  of the approximation (\ref{2.36})--(\ref{2.42}), however,
is that it yields a cavity function that has a jump at $r=1$, i.e.,
$y(1^+)\neq y(1^-)$, while the exact $y(r)$ and its first derivative
must be continuous at $r=1$.

The failure to satisfy the condition $y(1^+)=y(1^-)$ can be traced
back to the constraint (\ref{2.38}), which is a consequence of
imposing the physical requirements (\ref{17}) on the approximation
(\ref{9})--(\ref{2.34}) for non-nearest-neighbor interactions.
Nevertheless, the form (\ref{2.39}) satisfies Eq.\ (\ref{17}) for
arbitrary $\xi$, provided $\xi'$ is still defined by Eq.\
(\ref{2.40}). This strongly suggests to maintain Eq.\ (\ref{2.39}),
except that now the parameter $\xi$ is not tied to satisfy Eq.\
(\ref{2.38}) but  is instead  determined from the condition
$y(1^+)=y(1^-)$. Since this is not enough to guarantee that
$y'(1^+)=y'(1^-)$, we complement the above approximation by adding a
linear function in the shell $0<r<1$. In summary, our
\textit{low-temperature} (LT) approximation consists of
\beq
g(r)=(r-1)A\Theta(1-r)+\sum_{n=0}^\infty \psi_n(r-n)\Theta(r-n),
\label{4.1}
\eeq
where the functions $\psi_n(r)$ are given by Eq.\ (\ref{21}), $\xi'$
being defined by Eq.\ (\ref{2.40}). The LT approximation (\ref{4.1})
contains two free parameters ($\xi$ and $A$) to be determined, as
\kk{indicated} before, from the continuity conditions of $y(r)$ at
$r=1$. To that end, note that
\beq
y(r<1)=\frac{A}{1-x}(r-1)+\frac{\xi-\xi'}{\rho(1-x)}e^{-\xi' r},
\label{4.2}
\eeq
\beqa
y(1<r<2)&=&\frac{\xi-\xi'}{\rho}e^{-\xi'
r}+\frac{\xi'}{\rho}\left[1+(\xi-\xi')(r-1)\right]\nn &&\times
e^{-\xi'(r-1)}.
\label{4.3}
\eeqa
Now, the condition $y(1^+)=y(1^-)$ yields the following
transcendental equation for the parameter $\xi$:
\beq
\frac{x}{1-x}(\xi-\xi')e^{-\xi'}=\xi',
\label{4.4}
\eeq
which  differs from Eq.\ (\ref{2.38}) in the replacement of
$e^{-\xi}$ by $e^{-\xi'}$. Finally, the condition $y'(1^+)=y'(1^-)$
determines the parameter $A$ as
\beq
A=\frac{1-x}{\rho}\xi'(\xi-\xi').
\label{4.5}
\eeq
This closes the construction of the LT approximation. \kk{It is
given by Eq.\ (\ref{4.1}), supplemented by Eqs.\ (\ref{2.40}),
(\ref{21}), (\ref{4.4}), and (\ref{4.5}).}

The Laplace transform of Eq.\ (\ref{4.1}) is
\beq
G(t)=A\frac{1-t-e^{-t}}{t^2}+ \frac{1}{\rho}\frac{P_1(t)}{1-
P_1(t)},
\label{4.6}
\eeq
where the function $P_1(t)$ is given by Eq.\ (\ref{2.39}) but now it
cannot be interpreted simply as the Laplace transform of the
nearest-neighbor probability distribution. As already mentioned, the
LT approximation (\ref{4.6}) fulfills the physical condition
(\ref{17}).

\begin{table}[htb]
\begin{ruledtabular}
\begin{tabular}{ccc}
$r$& $y(r)$ & $y'(r)$\\
\hline
0&$\xi' e^{\xi'}[1-(1-x)\xi']/\rho x$ &$-{\xi'}^2e^{\xi'}/\rho$ \\
1&$\xi'/\rho x$ &$-{\xi'}^2[1-(1-x)e^{\xi'}]/\rho x$\\
$2^-$&$\xi'[(1-x)\xi'+e^{-\xi'}]/\rho
x$&$-{\xi'}^2[e^{-\xi'}-(1-x)(1-\xi')]/\rho x$\\
$2^+$&$\xi'[(1-x)\xi'+e^{-\xi'}]/\rho
x$&${\xi'}^2[1-(1-x)\xi'-e^{-\xi'}]/\rho x$
\end{tabular}
\caption{Values of the cavity function and its first derivative at
$r=0$, 1, and 2 in the LT approximation (\protect\ref{4.1}).}
\label{y0}
\end{ruledtabular}
\end{table}
As an application of the explicit character of the LT approximation
(\ref{4.1}), we give in Table \ref{y0} the values of the cavity
function and its first derivative at $r=0$, 1, and 2. {}By
eliminating the parameter $\xi'$ in favor of $y(1)$ as $\xi'=\rho
xy(1)$ one gets
\beq
y(0)=y(1) e^{\rho x y(1)}\left[1-\rho x (1-x)y(1)\right],
\label{zsep1}
\eeq
\beq
\left.\frac{\partial \ln y(r)}{\partial r}\right|_{r=0}=-\frac{\rho
x^2 y(1)}{1-\rho x(1-x) y(1)},
\label{zsep2}
\eeq
\beq
y'(2^+)-y'(2^-)=\rho x^2[y(1)]^2.
\label{yp2}
\eeq
 Equations (\ref{zsep1})
and (\ref{zsep2}) can be considered as (approximate) extensions to
finite temperatures ($x\neq 1$) of the zero separation theorems
(\ref{39}) and (\ref{37}), respectively. Equation (\ref{yp2}) gives
the discontinuity of the slope of $g(r)$ at $r=2$. It is interesting
to note that, although Eq.\ (\ref{yp2}) has been derived from our LT
approximation, it is not only exact in the limit $x\to 1$, but also
in the opposite limit $x\to 0$ (at $\rho x=\text{const}$), as can be
verified by comparison with Eq.\ (\ref{d2}).

\subsection{Temperature expansion}

While the approximation (\ref{4.1}) is expected to become more and
more reliable as the temperature decreases, it is not restricted
\textit{a priori} to low temperatures since it incorporates all the
orders in the parameter $\overline{x}\equiv 1-x=e^{-1/T^*}$, i.e.,
\beq
\psi_n(r)=\psi_n^{(0)}(r)+\psi_n^{(1)}(r)\overline{x}+\psi_n^{(2)}(r)\overline{x}^2+\cdots.
\label{4.7}
\eeq
Let us now obtain explicitly $\psi_n^{(0)}(r)$ and
$\psi_n^{(1)}(r)$. In the limit $\overline{x}\to 0$, the solution of
Eq.\ (\ref{4.4}) coincides with that of Eq.\ (\ref{2.38}), namely
$\xi=\xi_0-\xi_0^2 e^{\xi_0}\overline{x}+\cdots$ and $\xi'=\xi-\xi_0
e^{\xi_0}\overline{x}+\cdots$, where $\xi_0$ is defined by Eq.\
(\ref{2.43}). Therefore, from Eq.\ (\ref{21}) we get
\beq
\psi_0^{(0)}(r)=0,\quad \psi_n^{(0)}(r)=\frac{\xi_0^n}{\rho
(n-1)!}e^{-\xi_0 r}r^{n-1},\quad n\geq 1,
\label{4.8}
\eeq
\beq
\psi_n^{(1)}(r)=\frac{\xi_0^{n+1}}{\rho^2 n!}e^{-\xi_0(
r-1)}r^{n-1}\left[\xi_0(n+1-\rho)r-n^2\right].
\label{4.9}
\eeq
{}From $\psi_0^{(1)}(r)$ and $\psi_n^{(0)}(r)$ for $n\geq 1$, and
taking into account that Eq.\ (\ref{4.5}) yields
$A=\xi_0^2e^{\xi_0}\overline{x}^2/\rho+\cdots\sim \overline{x}^2$,
one recovers the exact hard-rod result (\ref{2.46}).

\subsection{Density expansion}
In the low density regime, the solution of Eq.\ (\ref{4.4}) is
\beq
\xi=\rho\left(1+x\rho+x^3\rho^2\right)+\mathcal{O}(\rho^4),
\label{28}
\eeq
\beq
\xi'=x\rho\left(1+x^2\rho\right)+\mathcal{O}(\rho^3).
\label{30}
\eeq
Insertion into Eq.\ (\ref{4.5}) yields
\beq
A=x(1-x)^2\rho+\mathcal{O}(\rho^2).
\label{31}
\eeq
The density expansion of the functions $\psi_n(r)$ defined by Eq.\
(\ref{21}) is
\beq
\psi_0(r)=(1-x)\left[1+x\rho(1+x-r)\right]+\mathcal{O}(\rho^2),
\label{32}
\eeq
\beq
\psi_1(r)=x\left[1+\rho(x^2+r-2x r)\right]+\mathcal{O}(\rho^2),
\label{33}
\eeq
\beq
\psi_2(r)=x^2\rho r+\mathcal{O}(\rho^2),
\label{34}
\eeq
\beq
\psi_n(r)=\mathcal{O}(\rho^{n-1}),\quad n\geq 3.
\label{35}
\eeq
Using (\ref{31})--(\ref{34}) in the LT approximation (\ref{4.1}) one
recovers the exact result $y(r)=1+\rho y_1(r)+\mathcal{O}(\rho^2)$,
where $y_1(r)$ is shown in Table \ref{Table2}. \kk{Note that the}
inclusion of the term in Eq.\ (\ref{4.1}) headed by the parameter
$A$ is essential to get this result. However, Eq.\ (\ref{4.1}) does
not reproduce the exact $y_2(r)$ for $r<2$, except, of course, at
$x=1$.

\subsection{Test for a very low temperature}

\begin{figure}[htb]
\includegraphics[width=\columnwidth]{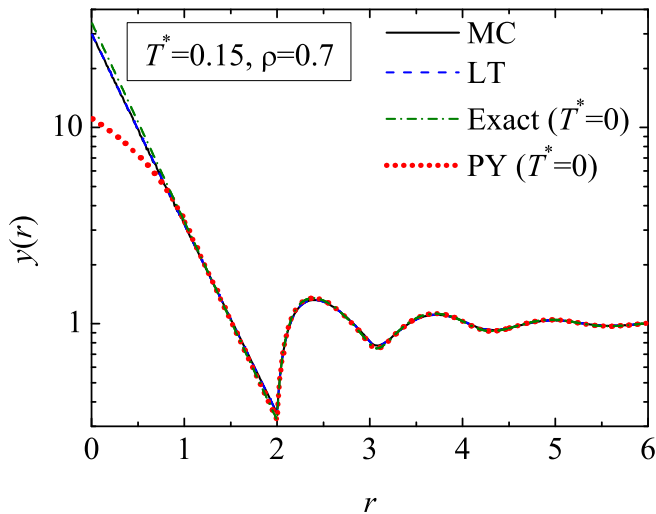}
\caption{(Color online) Plot of the cavity function $y(r)$ at
$\rho=0.7$ and $T^*=0.15$. The solid line correspond to MC
simulations and the dashed line (which  is practically
indistinguishable from the solid line) represents the
low-temperature (LT) approximation (\ref{4.1}). The figure also
includes the exact (dashed-dotted line) and the PY (dotted line)
cavity functions at zero temperature.
\label{yT015}}
\end{figure}
As a test of the reliability of the LT approximation (\ref{4.1}) for
very low temperatures, we present in Fig.\ \ref{yT015} a comparison
with MC simulations at $\rho=0.7$ and $T^*=0.15$. The agreement is
excellent. In fact, the simulation data cannot be distinguished from
the theoretical results. While the temperature $T^*=0.15$ is very
low ($\overline{x}=1-x\simeq 1.3\times 10^{-3}$), one can observe
that $y(r)$ in the region $r<1$ is slightly smaller than the one at
$T^*=0$. As in Fig.\ \ref{yHS}, Fig.\ \ref{yT015} shows the failure
of the PY approximation to account reasonably well for the cavity
function inside the core. As a bonus, Fig.\ \ref{yT015} illustrates
that MC simulations of $g(r)$ for penetrable spheres at low
temperatures (say $T^*\lesssim 0.2$) can be used to get (by
extrapolation) the cavity function of \textit{hard spheres} inside
the core. This quantity  is  accessible from simulations of true
hard spheres by an alternative method.\cite{LM84}

\section{Comparison of Monte Carlo simulations with theoretical approximations\label{sec5}}

As Figs.\ \ref{yT5&10} and \ref{yT015} illustrate, the HT
approximation (\ref{3.4}) and the LT approximation (\ref{4.1})
become very accurate in their expected domains of sufficiently high
and low temperatures, respectively. However, the physically most
interesting cases correspond to moderate temperatures (say
$0.5\lesssim T^*\lesssim 2$), where the finite penetrability of the
\kk{particles} plays a relevant role. Now \kk{some} interesting
questions are as follows: Up to what temperature the LT
approximation remains reasonably accurate? Below which temperature
the HT approximation ceases to be reliable? Do the answers to those
questions depend on the density? To address these points we have
performed MC simulations of the system\cite{note} for four
temperatures ($T^*=0.3$, 0.8, 1.5, and 3.0) and, in each case, for
three densities which roughly correspond to $\rho x\simeq 0.3$,
$0.5$, and $0.7$, except for the highest temperature ($T^*=3.0$), in
which case we have taken $\rho x\simeq 0.5$, $0.8$, and $1.1$.

\begin{figure}[htb]
\includegraphics[width=0.96\columnwidth]{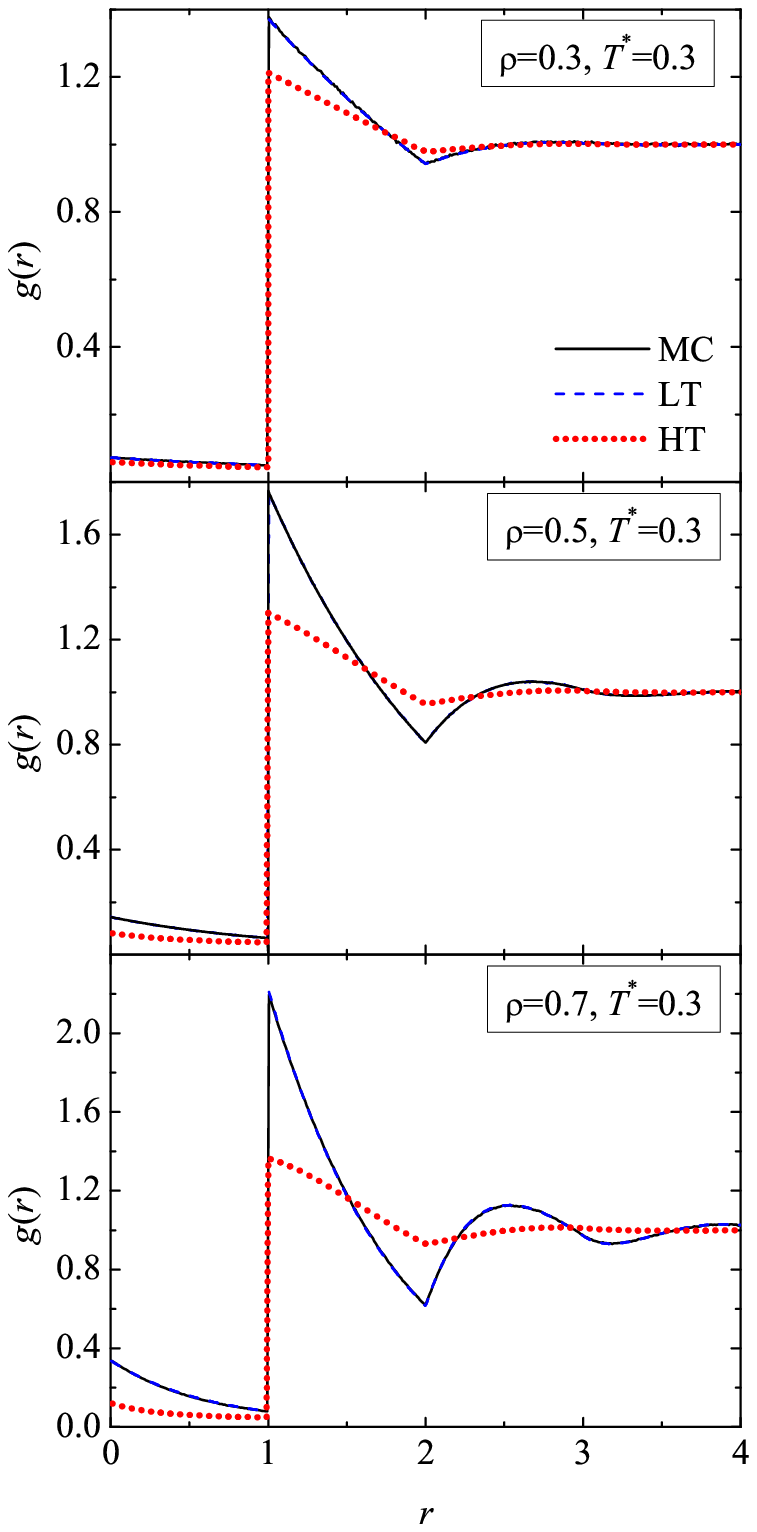}
\caption{(Color online) Plot of the radial distribution function
$g(r)$ at  $T^*=0.3$ and $\rho=0.3$ (top panel), $\rho=0.5$ (middle
panel), and $\rho=0.7$ (bottom panel). The solid lines correspond to
MC simulations,  the dashed lines (which  are practically
indistinguishable from the solid lines) represent the
low-temperature (LT) approximation (\ref{4.1}), and the dotted lines
represent the high-temperature (HT) approximation (\ref{3.4}).
\label{gT03}}
\end{figure}
Figure \ref{gT03} shows the simulation data and the theoretical
predictions of the radial distribution function $g(r)$ at
temperature $T^*=0.3$ and densities $\rho=0.3$, $0.5$, and $0.7$.
Although this is a rather low temperature, overlapping effects start
to be important, especially as the density increases. It is then
remarkable that the LT approximation yields results that are
indistinguishable from the simulation data. On the one hand, as one
could have anticipated, the HT approximation does not capture the
rich structural features that are present at this relatively small
temperature. \kk{On the other hand, it is noteworthy that, even at a
temperature as low as $T^*=0.3$, the results provided by the HT
approximation are not unphysical.}

\begin{figure}[htb]
\includegraphics[width=0.96\columnwidth]{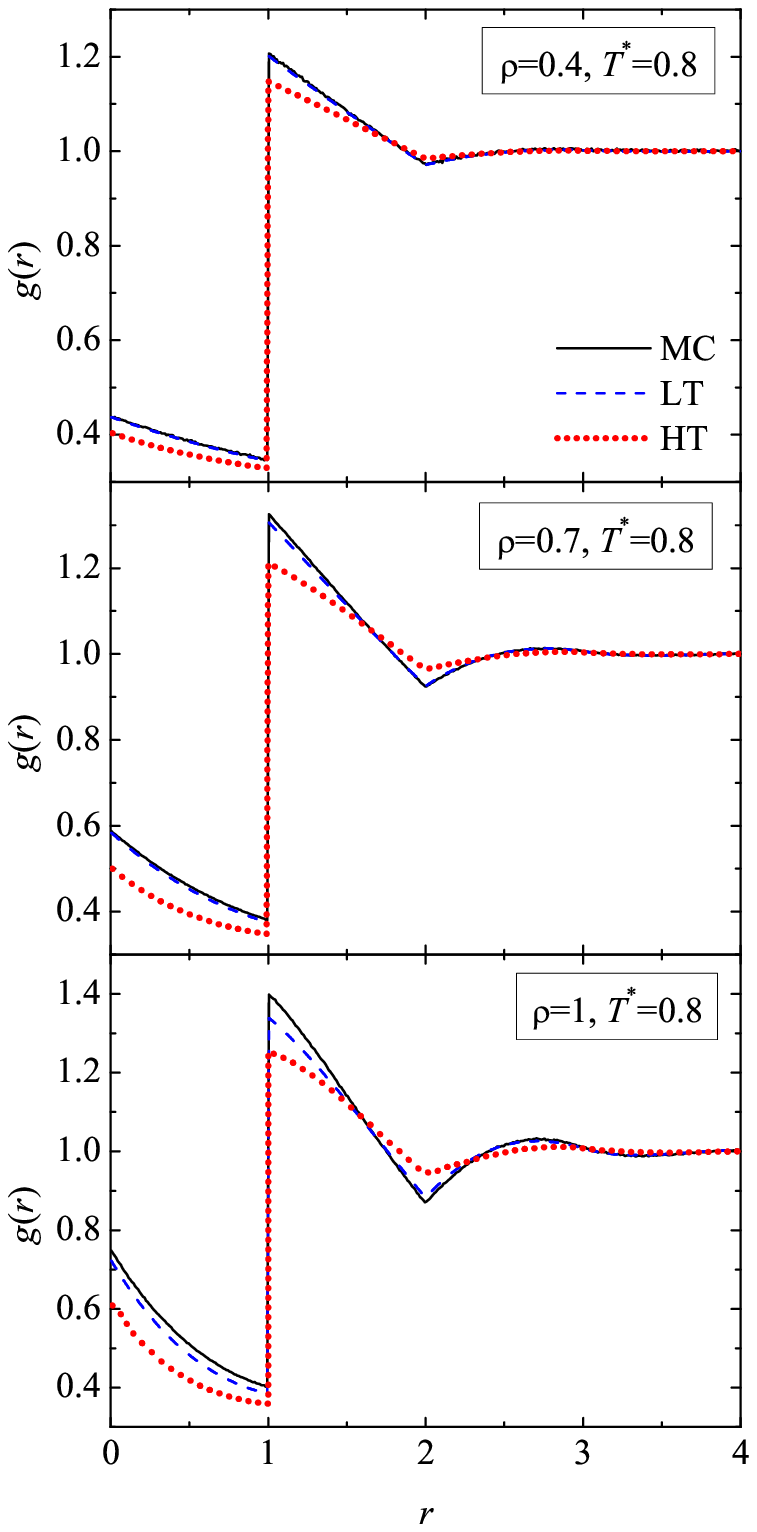}
\caption{(Color online) Plot of the radial distribution function
$g(r)$ at  $T^*=0.8$ and $\rho=0.4$ (top panel), $\rho=0.7$ (middle
panel), and $\rho=1$ (bottom panel). The solid lines correspond to
MC simulations,  the dashed lines (which  are practically
indistinguishable from the solid lines in the top and middle panels)
represent the low-temperature (LT) approximation (\ref{4.1}), and
the dotted lines represent the high-temperature (HT) approximation
(\ref{3.4}).
\label{gT08}}
\end{figure}
Next we consider in Fig.\ \ref{gT08} the moderate temperature
$T^*=0.8$ and the densities $\rho=0.4$, $0.7$, and $1$. The LT
approximation still yields excellent results, although it tends to
slightly underestimate $g(r)$ near contact and in the overlapping
region as density increases. The HT approximation is reasonable at a
qualitative level only.

\begin{figure}[htb]
\includegraphics[width=0.96\columnwidth]{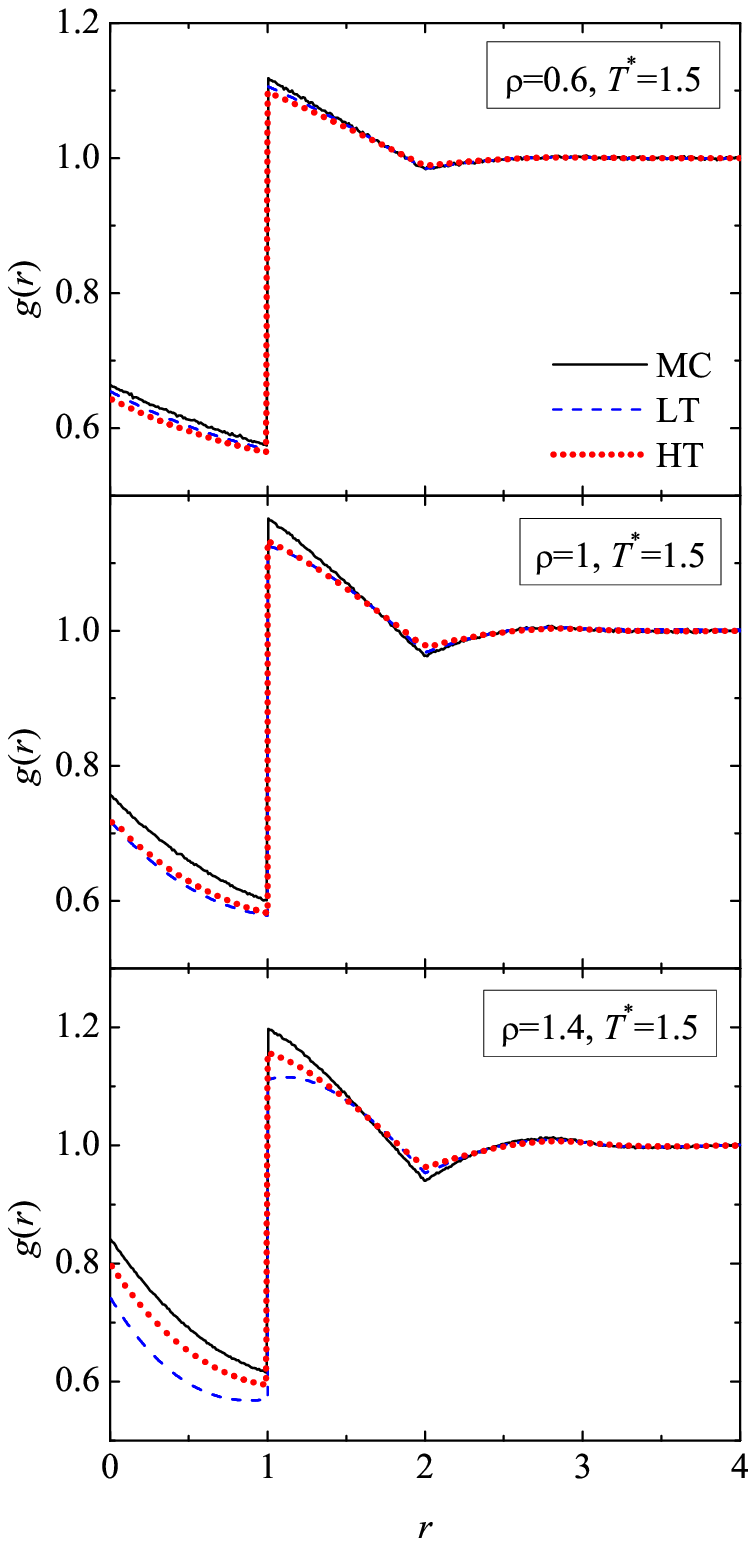}
\caption{(Color online) Plot of the radial distribution function
$g(r)$ at  $T^*=1.5$ and $\rho=0.6$ (top panel), $\rho=1$ (middle
panel), and $\rho=1.4$ (bottom panel). The solid lines correspond to
MC simulations,  the dashed lines represent the low-temperature (LT)
approximation (\ref{4.1}), and the dotted lines represent the
high-temperature (HT) approximation (\ref{3.4}). \kk{Note that the
dashed and dashed-dotted lines  in the middle panel are hardly
distinguishable.}
\label{gT15}}
\end{figure}
The temperature $T^*=1.5$ is a representative example of a
transitional value, as illustrated in Fig.\ \ref{gT15} for
$\rho=0.6$, $1$, and $1.4$. At the lowest density ($\rho=0.6$) the
LT theory is still quite good, the HT theory being only slightly
less accurate. At the intermediate density ($\rho=1$), both theories
practically coincide and underestimate $g(r)$ near contact and in
the overlapping region. Finally, at the highest density
($\rho=1.4$), the preferable theory is the HT one, except for
$r\gtrsim 1.5$. \kk{Thus, although the temperature is the most
important parameter to determine which approximation is better, the
density  can also play a relevant role.}

\begin{figure}[htb]
\includegraphics[width=0.96\columnwidth]{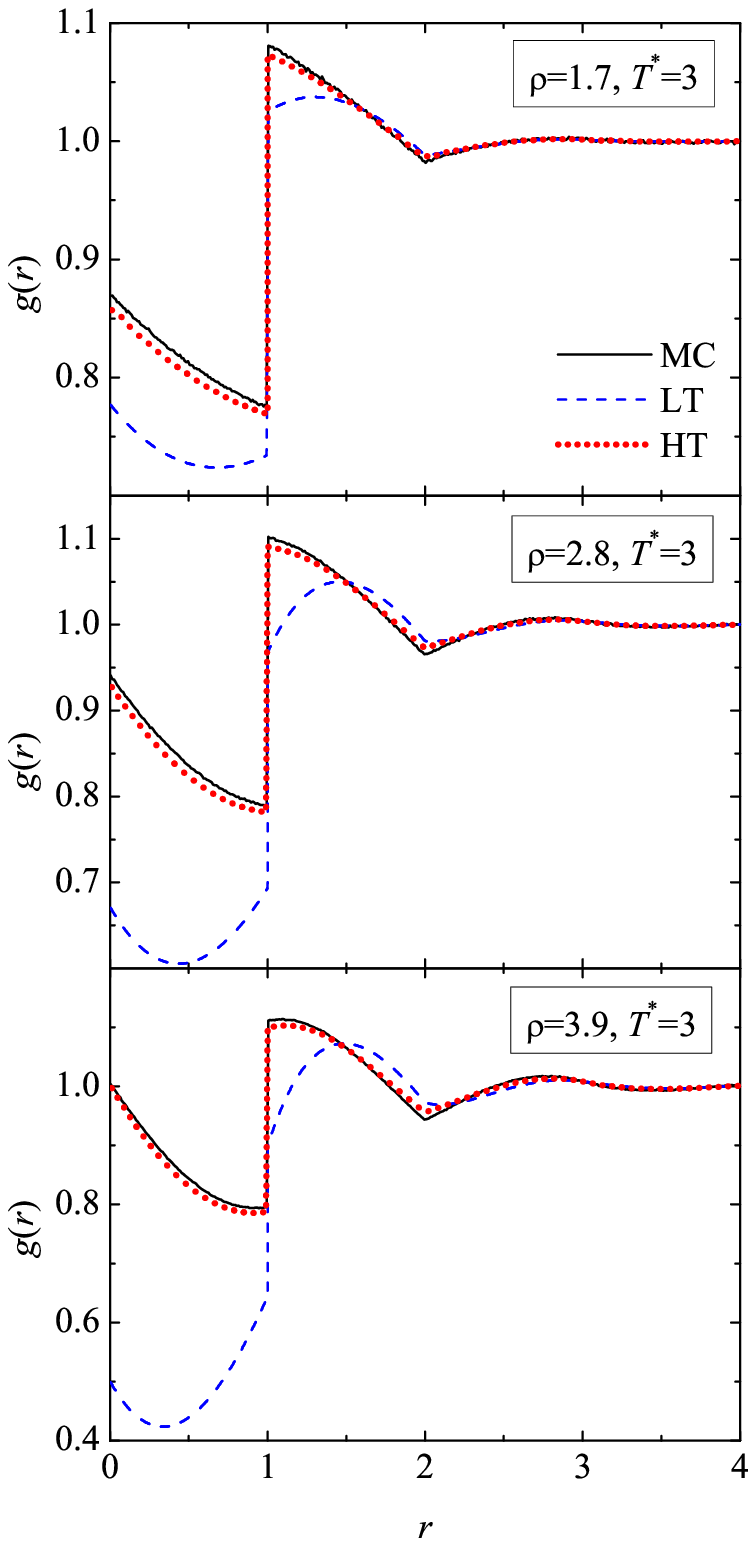}
\caption{(Color online) Plot of the radial distribution function
$g(r)$ at  $T^*=3$ and $\rho=1.7$ (top panel), $\rho=2.8$ (middle
panel), and $\rho=3.9$ (bottom panel). The solid lines correspond to
MC simulations,  the dashed lines represent the low-temperature (LT)
approximation (\ref{4.1}), and the dotted lines represent the
high-temperature (HT) approximation (\ref{3.4}).
\label{gT3}}
\end{figure}
Figure \ref{gT3} shows $g(r)$ at a higher temperature $T^*=3$ and
for the densities $\rho=1.7$, $2.8$, and $3.9$. Now the HT
approximation succeeds in reproducing the simulation data quite well
for the three densities, while the LT approximation becomes rather
poor, especially as the density increases.

It is interesting to note from the simulation curves in Figs.\
\ref{gT03}--\ref{gT3} that the curvature of $g(r)$ in the shell
$1<r<2$ changes from concave to convex as the temperature increases.
Moreover, at a given temperature, the magnitude of the curvature
increases with the density. \kk{We also observe that, as temperature
and density increase, so does the number of overlapped pairs.
Nevertheless, the radial distribution function in the
non-overlapping region ($r>1$) becomes less structured as
temperature increases.}

\begin{figure}[htb]
\includegraphics[width=\columnwidth]{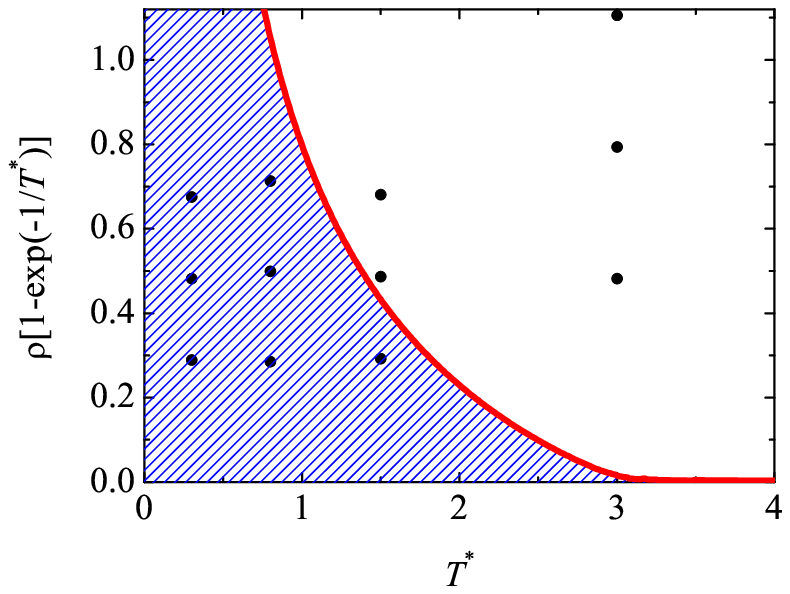}
\caption{(Color online) The thick solid line represents the locus of
points in the plane $\rho x$ vs $T^*$ where the LT and HT
approximations predict the same value of the contact quantity
$y(1)=g(1^+)$.  The LT approximation is accurate below the curve
(shaded region), while the HT approximation is accurate above the
curve. The circles represent the states considered in Figs.\
\ref{gT03}--\ref{gT3}.
\label{rhocrit}}
\end{figure}
The comparison between MC data and the LT and HT theories carried
out in Figs.\ \ref{gT03}--\ref{gT3} shows that both theories keep
being reasonably accurate well beyond their respective expected
domains. In general, the smaller the temperature and/or the density,
the better the LT theory, while the opposite happens for the HT
theory. Both theories complement each other so well that they meet
and become practically equivalent for intermediate temperatures and
densities, as illustrated by the middle panel of Fig.\ \ref{gT15}.
In order to define the ``basins'' of both theories in the
density-temperature plane, let us consider the locus of points
$\widetilde{\rho}(T^*)$ where the contact quantity $y(1)=g(1^+)$
takes the same value in both approximations. This locus is plotted
in Fig.\ \ref{rhocrit} in the representation $\rho x$ vs $T^*$. We
have used the scaled density $\rho x$ rather than the true density
$\rho$ because the range of values of $\rho$ accessible to the fluid
phase increases with increasing temperature roughly as $\rho\sim
1/x$.\cite{AS04} {}From Fig.\ \ref{rhocrit} and the comparison with
MC results carried out in Figs.\ \ref{yT5&10}--\ref{gT3} one can
conclude that the LT approximation is reliable for $T^*\lesssim 0.8$
at any density, while the HT approximation is reliable for
$T^*\gtrsim 3$ at any density. Between these two temperatures the
locus defined by the condition $y_{\text{LT}}(1)=y_{\text{HT}}(1)$
separates the basins of each approximation. In that transitional
regime, the smaller (larger) the density the better the LT (HT)
theory is. For states lying on the locus both approximations yield
practically equivalent predictions, which slightly underestimate
$g(r)$ for distances $r\lesssim 1.5$ (cf.\ Fig.\ \ref{gT15}). As one
departs from the locus, the quality of the corresponding
approximation (either LT or HT) significantly improves, as
illustrated by Figs.\ \ref{yT015}--\ref{gT08} in the LT case, and by
Figs. \ref{yT5&10} and  \ref{gT3} in the HT case. A global
approximation can be proposed just by adopting either the LT or the
HT approximation, depending on which basin the state point lies  in.
More specifically,
\beqa
g(r;\rho,T^*)&=&\Theta(\alpha(\rho,T^*))g_{\text{LT}}(r;\rho,T^*)\nn
&&+\Theta(-\alpha(\rho,T^*))g_{\text{HT}}(r;\rho,T^*),
\label{5.1}
\eeqa
 where
\beq
\alpha(\rho,T^*)=y_{\text{LT}}(1;\rho,T^*)-y_{\text{HT}}(1;\rho,T^*).
\label{5.2}
\eeq

\begin{figure}[htb]
\includegraphics[width=0.96\columnwidth]{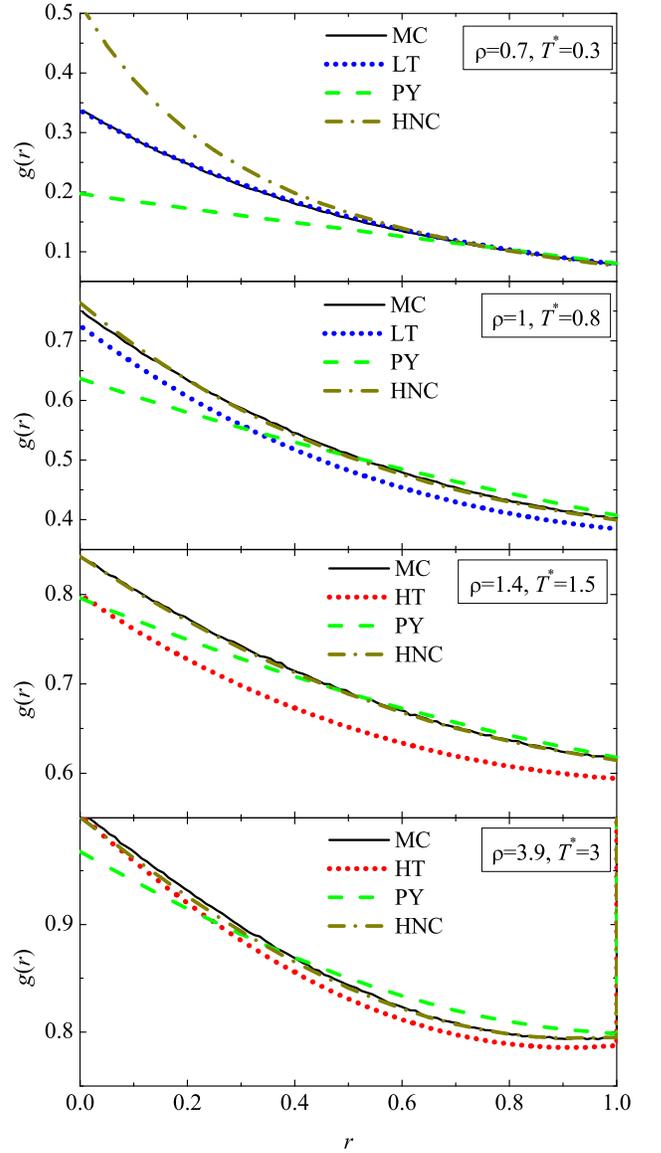}
\caption{(Color online) Radial distribution function inside the core
at four different states, as obtained from MC simulations (solid
lines), the LT or HT approximations (dotted lines), the PY theory
(dashed lines), and the HNC theory (dashed-dotted lines). Note that
the LT results are indistinguishable from the MC data in the top
panel, while the HNC results are \kk{hardly distinguishable} from
the MC data in the remaining three panels.
\label{PYHNC}}
\end{figure}
While we have been concerned in this paper with \kk{heuristic}
\textit{analytical} approximations, it is worthwhile comparing our
MC data with the two classical integral equation theories. For
one-dimensional systems, the PY and HNC integral equations  read
\beq
y(r)=1+\rho\int_{-\infty}^\infty \dd s\, h(|s|)y(|r-s|)f(|r-s|),
\label{5.3}
\eeq
\beq
\ln y(r)=\rho\int_{-\infty}^\infty \dd s\, h(|s|)\left[h(|r-s|)-\ln
y(|r-s|)\right],
\label{5.4}
\eeq
respectively, where we recall that
$h(r)=e^{-\varphi(r)/k_BT}y(r)-1$. We have numerically solved the
above two equations by a standard iterative method for the simulated
states. The agreement is always quite good outside  the core
($r>1$), so that we focus now on the overlapping region. Figure
\ref{PYHNC} shows $g(r)$ with $r<1$ for the  temperatures $T^*=0.3$,
$0.8$, $1.5$, and $3$, and, in each case, for the highest density
considered in the simulations. Apart from the simulation data and
the PY and HNC predictions, we have also included either our LT
approximation ($T^*=0.3$ and $0.8$) or our HT approximation
($T^*=1.5$ and $3$), in accordance with Fig.\ \ref{rhocrit}. At the
lowest temperature ($T^*=0.3$) the PY and HNC predictions seriously
underestimate and overestimate, respectively, the simulation data,
while the LT approximation is excellent. However, at moderate and
high temperatures ($T^*=0.8$,  $1.5$, and $3$) the HNC approximation
becomes very good.  The PY approximation is still worse than our LT
approximation at $T^*=0.8$ \kk{but}  is more accurate than our HT
approximation at $T^*=1.5$. \kk{At $T^*=3$, the PY and HT
approximations exhibit a comparable accuracy}. Note that as
temperature increases, the HT, PY, and HNC approximations tend to
coincide and reduce to the exact asymptotic behavior (\ref{2}). The
good performance of the HNC approximation at moderate temperatures
reported here contrasts with the situation reported in the
three-dimensional case.\cite{FLL00,RSWL00}

\section{Summary and discussion\label{sec6}}
\kk{It is obvious that many of the physical properties of real 3D
systems cannot be fully represented by 1D models. For instance,
entropy-driven phase transitions, such as the fluid-solid transition
in hard spheres or the demixing transition in positively nonadditive
hard-sphere mixtures, are absent in 1D. In the case of penetrable
spheres, however, the 1D model retains a number of features of its
2D and 3D counterparts (such as the lack of an exact solution, the
formation of ``clumps'' of overlapped particles,\cite{KGRCM94} or
the existence of a fluid-solid phase transition,\cite{AS04}) what
justifies its study, apart from its interest at a more fundamental
level.}

In this paper we have investigated the structural properties of
one-dimensional fluids of penetrable spheres. First we have derived
the exact expression of the radial distribution function $g(r)$
through second order in density at any temperature, the results
being presented in Table \ref{Table2}. This calculation already
reveals some of the main features of the PY and HNC approximations:
the former underestimates $g(r)$ inside the core ($r<1$), while the
latter overestimate it. Moreover, the HNC predictions become better
than the PY ones for temperatures larger than $T^*\approx 1$.

In the limit of asymptotically high temperatures only the linear
chain diagrams contributing to the virial expansion of the cavity
function need to be retained. The series can then be resummed, the
result being given by Eqs.\ (\ref{2}) and (\ref{4}). The auxiliary
function $w(r)$, which depends on density and temperature \kk{only}
through the scaled quantity $\rho x$, where $x$ is defined by Eq.\
(\ref{2.18}), embodies the structural properties of the
high-temperature penetrable-rod fluid. Its series in powers of $\rho
x$ converges uniformly for $\rho x<\frac{1}{2}$, so that this is the
radius of convergence of the virial series of the equation of state.
An explicit representation of $w(r)$ for each shell $n<r<n+1$ has
been derived, Eq.\ (\ref{2.26}), what has allowed us to get some
simple relations, Eqs.\ (\ref{d1})--(\ref{d6}).

The high-temperature limit has been complemented by the
zero-temperature limit. In that limit the system becomes a hard-rod
fluid, whose expression for $g(r)$ is known exactly. Since the
particles are impenetrable in this limit, $g(r)$ vanishes for $r<1$.
However, the cavity function $y(r)$ remains finite and its
determination is important to understand the behavior of $g(r)$ for
low (but non-zero) temperatures. In order to derive $y(r)$ for hard
rods, we have first applied to finite temperatures the scheme
providing the exact solution in the case of interactions restricted
to nearest neighbors. Although the penetrable-rod potential at
non-zero temperature allows for multiple interactions, the
nearest-neighbor approximation becomes more and more reliable as the
temperature decreases, providing thus the exact $y(r)$ for $r<1$ in
the hard-rod limit. We are not aware of any previous derivation of
$y(r)$ for hard rods inside the core, although it might possibly
exist. Quite surprisingly, the expression of $y(r)$ for $r<1$ is
just the analytical continuation of its expression for $1<r<2$, so
that the hard-rod cavity function is fully analytical at $r=1$, even
though the potential is \kk{highly} singular at that point. It is
well known that the PY approximation gives the exact $g(r)$ for hard
rods due a fortunate mutual cancelation of the neglected diagrams in
the region $r>1$. However, it is perhaps less known\cite{M69} that
such a cancelation does not take place for $r<1$, what results in a
wrong PY prediction of $y(r)$ inside the core, already to second
order in density. In fact, the PY  and HNC cavity functions for hard
rods exhibit \kk{an artificial} second-order discontinuity at $r=1$.
The failure of the PY theory to reproduce the hard-rod $y(r)$ inside
the core is a precursor of its tendency to dramatically
underestimate the penetrability phenomenon at finite temperatures.

It seems tempting to exploit the exact asymptotic behavior of the
correlation functions for high temperatures by extrapolating it to
finite temperatures. Among several possibilities, we have taken the
Pad\'e approximant (\ref{3.4}) as our high-temperature (HT)
approximation. It  is equivalent to retaining, in addition to all
the linear chain diagrams, all the reducible diagrams factorizing
into linear chain diagrams, although those reducible diagrams have
enhanced factors that partially compensate for the neglected
irreducible diagrams. In addition, while $x\approx 1/T^*$ for high
temperatures, we have kept the nonlinear relationship (\ref{2.18})
in the extension to finite temperatures since  this is the parameter
naturally appearing in the virial expansion. Despite its simplicity,
we have checked by comparison with MC simulations that this HT
approximation does quite a good job for temperatures higher than
$T^*\approx 3$ at any density.

Obviously, the HT approximation is not accurate for low
temperatures. To complement it we have constructed a low-temperature
(LT) approximation, which is based on the explicit expression for
$g(r)$ obtained from the nearest-neighbor approximation mentioned
above but introduces two changes. First, the transcendental equation
that the damping coefficient $\xi$ obeys in the nearest-neighbor
approximation is replaced by another one coming from the continuity
condition of $y(r)$ at $r=1$. Next, a linear function is added in
the region $r<1$ to enforce continuity of the first derivative of
$y(r)$ at $r=1$. More specifically, in the LT approximation $g(r)$
is given by Eq.\ (\ref{4.1}), where the functions $\psi_n(r)$ are
defined by Eq.\ (\ref{21}) and the amplitude $A$ is given by Eq.\
(\ref{4.5}); in those equations, the parameter $\xi'$ is defined by
Eq.\ (\ref{2.40}) and finally the parameter $\xi$ is the solution to
the transcendental equation (\ref{4.4}). This LT approximation
reduces to the exact solution in the hard-rod limit. In addition, it
is excellent for low temperatures (e.g., $T^*=0.3$) and  quite
 good  up to $T^*\approx 0.8$ at any density.

 For intermediate temperatures ($0.8\lesssim T^*\lesssim 3$), the LT
 and HT theories compete each other, the former being better for densities
lower than a certain threshold value $\widetilde{\rho}(T^*)$ and the
latter  being better for $\rho>\widetilde{\rho}(T^*)$. In fact, both
theories yield almost identical predictions for states lying on the
curve $\widetilde{\rho}(T^*)$. As one departs from that curve by
decreasing (increasing) the density and/or temperature the quality
of the LT (HT) approximation  improves significantly. Therefore,
given the simplicity and analytic character of both approximations,
it is quite reinforcing that they match and complement so well that
their combined use [cf.\ Eqs.\ (\ref{5.1}) and (\ref{5.2})] covers
satisfactorily the whole range of densities and temperatures. Of
course, it would be much nicer to have a unique analytic
approximation being equally accurate for both low and high
temperatures. However, we have not been able to devise  such a
unique approach.

To put these results in a proper perspective, we have also compared
our MC data with numerical solutions of the PY and HNC integral
equations. The results show that the PY approximation is quite poor
at low temperatures, slowly improving as the temperature increases.
The HNC approximation, on the other hand, is inaccurate only at low
temperatures, being excellent at moderate and high temperatures
($T^*\gtrsim 0.8$).

Although restricted to one-dimensional systems,  the research
carried out in this paper can be used to pave the way to similar
approaches in the more realistic case of three-dimensional
penetrable-sphere fluids. Since the asymptotic high-temperature
behavior is given by Eqs.\ (\ref{2}) and (\ref{4.1b}) for any
dimensionality,\cite{AS04} the proposal of the HT approximation is
straightforward. The only difference is that now the
(three-dimensional) Fourier transform of the auxiliary function
$w(r)$ is given by
\beq
{\wk}(k)=96\pi {\eta}x\frac{(k\cos k-\sin k)^2}
{k^3\left[k^3-24{\eta}x\left(k\cos k-\sin k\right)\right]},
\label{6.1}
\eeq
where $\eta=(\pi/6)\rho$ is the packing fraction. The construction
of the LT approximation is less direct since the exact solution in
the hard-sphere limit is not known. However, we can use the analytic
solution of the PY equation for hard spheres\cite{B74,HM86,WT63} as
a starting point. The structure of the resulting proposal is
\beq
g(r)=\widehat{g}(r)\exp\{(r-1)\left[A+B r(r+1)\right]\Theta(1-r)\},
\label{6.2}
\eeq
where the Laplace transform of $r\widehat{g}(r)$ is
\beq
\widehat{G}(t)=\frac{t}{12\eta} \frac{P(t)}{1+P(t)},\quad P(t)=
\frac{L_0-e^{-t}(L_0+1+L_1 t)}{1+S_1 t+S_2 t^2+S_3 t^3}.
\label{6.3}
\eeq
In Eqs.\ (\ref{6.2}) and (\ref{6.3}), the coefficients $A$, $B$,
$L_i$, and $S_i$ are functions of $\eta$ and $T^*$ determined by
imposing consistency conditions. In the limit $T^*\to 0$ the
coefficients $A$, $B$, and $L_0$ vanish and one recovers the radial
distribution function for hard spheres in the PY approximation.
However, the cavity function $y(r)$ in the region $r<1$ differs from
the PY one even in that limit, \kk{in parallel to what happens in
the 1D case.} Preliminary  comparisons of the HT and LT
approximations (\ref{6.1}) and (\ref{6.2}) with MC simulations are
quite encouraging. A detailed report will be published
elsewhere.\cite{MYS05}

\acknowledgments One of the authors (A.M.) is grateful to the
hospitality of the University of Extremadura and its Social Council
during a stay in May and June 2004, when this work was started. His
research has been
 partially supported by the Ministry of Education, Youth, and Sports of
the Czech Republic under the project No.\ LC 512. The research of
the other author (A.S.) has been supported by the Ministerio de
Educaci\'on y Ciencia (Spain) through Grant No.\ FIS2004-01399
(partially financed by FEDER funds) and by the European Community's
Human Potential Programme under contract No.\ HPRN-CT-2002-00307,
DYGLAGEMEM.

\appendix*
\section{Derivation of Equations (\protect\ref{2.25}), (\protect\ref{2.29}), and (\protect\ref{2.26})\label{appA}}

Expanding (\ref{4.1b}) in powers of density, one gets (\ref{2.24})
with
\beqa
w_n(r)&=&\frac{1}{2\pi}\int_{-\infty}^\infty \dd k\,
e^{ikr}\left[\fk_\hs(k)\right]^n\nn &=&
\frac{1}{2\pi}\int_{-\infty}^\infty \dd k\,
e^{ikr}\left(-\frac{2\sin k}{k}\right)^n\nn &=& \frac{i^n}{2\pi}
\sum_{m=0}^n \binom{n}{m}(-1)^{n+m}\int_L \dd k\,
\frac{e^{ik(r+2m-n)}}{k^n},\nn
\label{A1}
\eeqa
where $L$ is a path in the complex plane from $k=-\infty$ to
$k=+\infty$ that goes round the singularity at $k=0$ from above. If
$r+2m-n>0$, we can close the path with an upper half circle of
infinite radius, so that the integral identically vanishes. If
$r+2m-n<0$ we close the path with a lower half circle, so that the
contour encircles a pole at $k=0$ of order $n$. By applying the
residue theorem, we then have
\beqa
\int_L \dd k\,
\frac{e^{ik(r+2m-n)}}{k^n}&=&\frac{2\pi}{(n-1)!}(-i)^n
(n-2m-r)^{n-1}\nn &&\times\Theta(n-2m-r).
\label{A2}
\eeqa
Inserting this into Eq.\ (\ref{A1}) one gets Eq.\ (\ref{2.25}).

Let us now prove Eq.\ (\ref{2.29}). First, we note that, by
symmetry,
\beq
\frac{1}{2\pi i}\int_{-\infty}^\infty \dd k\,\frac{(-2\sin
k)^{n+1}}{k^n}=0.
\label{A7}
\eeq
By following on the left-hand side steps similar to those followed
to derive Eqs.\ (\ref{A1}) and (\ref{A2}), we get the identity (for
$n\geq 2$)
\beq
0=\sum_{m=0}^{[(n+1)/2]} \frac{(-1)^{n+m+1}}{m!(n+1-m)!}
(n+1-2m)^{n-1},
\label{A6}
\eeq
where $[a]$ denotes the integer part of $a$. Next, setting $r=1$ in
Eq.\ (\ref{2.25}) and making the change $m\to m+1$, we have
\beq
\frac{w_n(1)}{n}=\sum_{m=0}^{[(n+1)/2]}\frac{(-1)^{n+m+1}m}{m!(n+1-m)!}
(n+1-2m)^{n-1}.
\label{A5}
\eeq
Therefore,
\beqa
\frac{w_{n+1}(0)}{n+1}+2\frac{w_n(1)}{n}&=&
(n+1)\sum_{m=0}^{[(n+1)/2]} \frac{(-1)^{n+m+1}}{m!(n+1-m)!}\nn
&&\times (n+1-2m)^{n-1}.
\label{A8}
\eeqa
On account of Eq.\ (\ref{A6}), the right-hand side of Eq.\
(\ref{A8}) vanishes, what proves Eq.\ (\ref{2.29}).

Finally, we want to collect together all the terms contributing to
the same Heaviside function in Eq.\ (\ref{2.24}). We first define
the index $p=2m-n$ and note that if $p=1$ then $n=1+2q$ and $m=1+q$
with $q=1,2,3,\ldots$, while if $p\geq 2$ then $n=p+2q$ and $m=p+q$
with $q=0,1,2,\ldots$. Therefore,
\beqa
w(r)&=&\sum_{q=1}^\infty \frac{(-1)^{1+q}(1+2q)}{(1+q)!q!}(\rho
x)^{2q}(1-r)^{2q}\Theta(1-r)\nn &&+ \sum_{p=2}^\infty
\sum_{q=0}^\infty \frac{(-1)^{p+q}(p+2q)}{(p+q)!q!}(\rho
x)^{p+2q-1}(p-r)^{p+2q-1}\nn &&\times \Theta(p-r).
\label{A3}
\eeqa
Making use of the identity\cite{AS72}
\beq
\sum_{q=0}^\infty \frac{(-1)^q(p+2q)}{(p+q)! q!}
z^{p+2q-1}=2J'_{p}(2z),
\label{A4}
\eeq
where $J_p'=\frac{1}{2}(J_{p-1}-J_{p+1})$ is the first derivative of
the Bessel function of the first kind $J_p$, Eq.\ (\ref{A3}) reduces
to Eq.\ (\ref{2.26}).


\begin{thebibliography}{00}

\bibitem{L01}
C. N. Likos, Phys. Rep. \textbf{348},  267 (2001), and references
therein.

\bibitem{LLWAJAR98}
C. N. Likos, H. L\"owen, M. Watzlawek, B. Abbas, O. Jucknischke, J.
Allgaier, and D. Richter, Phys. Rev. Lett. \textbf{80},  4450
(1998); M. Watzlawek, C. N. Likos, and H. L\"owen, \textit{ibid.}
\textbf{82},  5289 (1999).

\bibitem{SS97}
F. H. Stillinger and D. K. Stillinger, Physica A \textbf{244}, 358
(1997).

\bibitem{GL98}
H. Graf and H. L\"owen, Phys. Rev. E \textbf{57},  5744 (1998).

\bibitem{LLWL00}
A. Lang, C. N. Likos, M. Watzlawek, and H. L\"owen, J. Phys.:
Condens. Matter \textbf{12},  5087(2000).

\bibitem{LBH00}
A. A. Louis, P. G. Bolhuis, and J.-P. Hansen, Phys. Rev. E
\textbf{62},  7961 (2000).

\bibitem{LLWL01}
C. N. Likos, A. Lang, M. Watzlawek, and H. L\"owen, Phys. Rev. E
\textbf{63},  031206 (2001).

\bibitem{FHL03}
R. Finken, J.-P. Hansen, and A. A. Louis, J. Stat. Phys.
\textbf{110},  1015 (2003).

\bibitem{MW89} C. Marquest and T. A.  Witten, J. Phys. (France) \textbf{50},  1267 (1989).

\bibitem{KGRCM94}
W. Klein, H. Gould, R. A. Ramos, I. Clejan, and A. I. Mel'cuk,
Physica A \textbf{205},  738 (1994).

\bibitem{LWL98} C. N. Likos, M. Watzlawek, and H. L\"owen, Phys. Rev.
E \textbf{58},  3135 (1998).

\bibitem{S99}
M. Schmidt, J. Phys.: Condens. Matter \textbf{11}, 10163 (1999).

\bibitem{FLL00} M. J. Fernaud, E. Lomba, and L. L. Lee, J. Chem. Phys. \textbf{112},  810 (2000).

\bibitem{RSWL00}
Y. Rosenfeld, M. Schmidt, M. Watzlawek, and H. L\"owen, Phys. Rev. E
\textbf{62},  5006 (2000).

\bibitem{SF02}
M. Schmidt and M. Fuchs, J. Chem. Phys. \textbf{117},  6308 (2002).

\bibitem{KS02}
S.-C. Kim and S.-Y. Suh, J. Chem. Phys. \textbf{117},   9880 (2002).

\bibitem{CG03}
N. Choudhury and S. K. Ghosh, J. Chem. Phys. \textbf{119}, 4827
(2003).

\bibitem{AS04}
L. Acedo and A. Santos, Phys. Lett. A \textbf{323}, 427 (2004).

\bibitem{S05}
A. Santos, ``Kinetic Theory of Soft Matter. The Penetrable-Sphere
Model,''   in \textit{Rarefied Gas Dynamics: 24th International
Symposium on Rarefied Gas Dynamics}, edited by M. Capitelli (AIP
Conference Proceedings \textbf{762}, 2005), pp.\ 276--281;
cond-mat/0501068.

\bibitem{GDL79}
L. Groome, J. W.  Dufty,  and M. J. Lindenfeld., {Phys. Rev. A}
\textbf{19}, 304 (1979).

\bibitem{CS02}
J. A. Cuesta and A. S\'anchez, J. Phys. A: Math. Gen. \textbf{35},
2373 (2002); J. Stat. Phys. \textbf{115}, 869 (2004).

\bibitem{M93}
\kk{D. C. Mattis, \textit{The Many-Body Problem: An Encyclopedia of
Exactly Solved Models in One Dimension} (World Scientific,
Singapore, 1993).}

\bibitem{JdC90}
\kk{See, for instance, G. Jannink and J. des Cloizeaux, J. Phys.:
Condens. Matter \textbf{2}, 1 (1990); J. des Cloizeaux and G.
Jannink, \textit{Polymers in Solution: Their Modelling and
Structure} (Clarendon Press, Oxford, 1991).}


\bibitem{V64}
\kk{L. Verlet, Physica \textbf{30}, 95 (1964).}

\bibitem{M69}
\kk{A. M\"unster, \textit{Statistical Thermodynamics}, vol.\ 1
(Springer--Verlag, Berlin, 1969), Chap.\ 5.11.}

\bibitem{B74}
R. Balescu, \textit{Equilibrium and Nonequilibrium Statistical
Mechanics} (Wiley, New York, 1974).


\bibitem{HM86}
J.-P.~Hansen and I. R. McDonald, \textit{Theory of Simple Liquids},
(Academic Press, London, 1986).

\bibitem{BH76}
J. A. Barker and D. Henderson, Rev. Mod. Phys. \textbf{48}, 587
(1976).

\bibitem{S05b}
A. Santos, J. Chem. Phys. \textbf{123}, 104102 (2005).

\bibitem{G75}
D. J. Gates, Physica A \textbf{81},  47 (1975).

\bibitem{GK77}
N. Grewe and W. Klein, J. Math. Phys. \textbf{18},  1729, 1735
(1977).

\bibitem{KG80}
 W. Klein and N. Grewe, J. Chem. Phys. \textbf{72},  5456 (1980).

\bibitem{KB81}
 W. Klein and A. C. Brown, J. Chem. Phys. \textbf{78},  6960 (1981).

\bibitem{AS72}
 \textit{Handbook of
Mathematical Functions}, edited by M. Abramowitz and I. A. Stegun
(Dover, New York, 1972).

\bibitem{SZK53}
Z. W. Salsburg, R. W. Zwanzig, and J. G. Kirkwood, J. Chem. Phys.
\textbf{21}, 1098 (1953).

\bibitem{PL54}
I. Prigogine and S. Lafleur, Bull. Classe Sci. Acad. Roy. Bel.
\textbf{263}, 484, 497 (1954).

\bibitem{K55}
R. Kikuchi, J. Chem. Phys. \textbf{23}, 2327 (1955).


\bibitem{LZ71}
J. Lebowitz and D. Zomick, J. Chem. Phys. \textbf{54}, 3335 (1971).

\bibitem{HC04}
M. Heying and D. Corti, Fluid Phase Equil. \textbf{220}, 85 (2004).


\bibitem{S63}
\kk{G. Stell, Physica \textbf{29}, 517 (1963).}

\bibitem{L95}
L. L. Lee, J. Chem. Phys. \textbf{103}, 9388 (1995); L. L. Lee, D.
Ghonasgi, and E. Lomba, \textit{ibid.} \textbf{104}, 8058 (1996); L.
L. Lee and A. Malijevsk\'y, \textit{ibid.} \textbf{114}, 7109
(2001).

\bibitem{LM84}
S. Lab\'{\i}k and A. Malijevsk\'y, Mol. Phys. \textbf{53}, 381
(1984).

\bibitem{note}
\kk{We have employed the conventional Metropolis algorithm on an NVT
ensemble with periodic boundary conditions. Typically, $N=10^4$
particles are used. In each realization, the correlation functions
are measured every 100 MC sweeps and this cycle is repeated $2\times
10^4$ times. The time-averaged quantities are further averaged over
ten independent realizations.}

\bibitem{WT63}M. S. Wertheim, Phys. Rev. Lett. \textbf{10},   321
(1963);  E. Thiele, J. Chem. Phys. \textbf{39},  474 (1963).

\bibitem{MYS05}
Al.\ Malijevsk\'y, S. B. Yuste, and A. Santos, unpublished.


\end{thebibliography}
\end{document}